\documentclass[prx,twocolumn,floatfix,10pt,superscriptaddress,longbibliography]{revtex4-1}
\usepackage[utf8]{inputenc}

\usepackage{natbib}
\usepackage{float}
\usepackage{graphicx}
\usepackage{latexsym}
\usepackage{amssymb}
\usepackage{amsmath}
\usepackage{amsfonts}
\usepackage{gensymb}
\usepackage{bm}
\usepackage{hyperref}
\usepackage{braket}
\usepackage{bbold}
\usepackage[caption=false]{subfig}
\usepackage[dvipsnames]{xcolor}
\usepackage[normalem]{ulem}
\usepackage{siunitx}
\usepackage[version=4]{mhchem}

\usepackage[dvipsnames]{xcolor}
\definecolor{pal0}{rgb}{0.8941, 0.102 , 0.1098}
\definecolor{pal1}{rgb}{0.2157, 0.4941, 0.7216}
\definecolor{pal2}{rgb}{0.302 , 0.6863, 0.2902}
\definecolor{pal3}{rgb}{0.5961, 0.3059, 0.6392}
\definecolor{pal4}{rgb}{1.    , 0.498 , 0.    }

\renewcommand{\Im}{\operatorname{Im}}

\newcommand{\br}{\mathbf{r}}
\newcommand{\bq}{\mathbf{q}}

\newcommand{\VTG}{V_{\mathrm{TG}}}
\newcommand{\VBG}{V_{\mathrm{BG}}}

\hypersetup{
  pdfauthor = {},
  pdftitle = {}
}

\begin{document}

\title{ 3D microwave imaging of a van der Waals heterostructure}

\author{ Leonard W. Cao*}
\affiliation{Department of Physics, University of California, San Diego, CA 92093, USA}

\author{ Chen Wu*}
\affiliation{Department of Physics, University of California, San Diego, CA 92093, USA}

\author{ Lingyuan Lyu}
\affiliation{Department of Physics, University of California, San Diego, CA 92093, USA}

\author{ Liam Cohen}
\affiliation{Department of Physics, University of California, Santa Barbara, CA 93106, USA}

\author{ Noah Samuelson}
\affiliation{Department of Physics, University of California, Santa Barbara, CA 93106, USA}

\author{ Ziying Yan}
\affiliation{Department of Physics, University of California, San Diego, CA 92093, USA}

\author{ Sneh Pancholi}
\affiliation{Department of Physics, University of California, San Diego, CA 92093, USA}

\author{ Kenji Watanabe}
\affiliation{Research Center for Electronic and Optical Materials, National Institute for Materials Science, Namiki 1-1, Tsukuba, Ibaraki 305-0044, Japan}

\author{ Takashi Taniguchi}
\affiliation{Research Center for Materials Nanoarchitectonics, National Institute for Materials Science, Namiki 1-1, Tsukuba, Ibaraki 305-0044, Japan}

\author{ Daniel E. Parker}
\affiliation{Department of Physics, University of California, San Diego, CA 92093, USA}

\author{ Andrea F. Young}
\affiliation{Department of Physics, University of California, Santa Barbara, CA 93106, USA}

\author{ Monica T. Allen}
\affiliation{Department of Physics, University of California, San Diego, CA 92093, USA}

\begin{abstract}
Van der Waals (vdW) heterostructures offer a tunable platform for the realization of emergent phenomena in layered electron systems.
While scanning probe microscopy techniques have proven useful for the characterization of surface states and 2D crystals, the subsurface imaging of quantum phenomena in multi-layer systems presents a significant challenge. In 3D heterostructures, states that occupy different planes can simultaneously contribute to the signal detected by the microscope probe, which complicates image analysis and interpretation. Here we present a quantum imaging technique that offers a glimpse into the third dimension by resolving states out of plane: it extracts the charge density landscape of \textit{individual} atomic planes inside a vdW heterostructure, layer by layer. As a proof-of-concept, we perform \textit{layer-resolved} imaging of quantum Hall states and charge disorder in double-layer graphene using milliKelvin microwave impedance microscopy.
Here the discrete energy spectrum of the top layer enables transmission of microwaves through gapped states, thus opening direct access to quantum phases in the subsurface layer. 
Resolving how charge is distributed out-of-plane offers a direct probe of interlayer screening, revealing signatures of negative quantum capacitance driven by many-body correlations. 
At the same time, we extract key features of the band structure and thermodynamics, including gap sizes and the local electronic compressibility.
Notably, by imaging the real-space charge distribution on different atomic planes beneath the surface, we shed light on the roles of surface impurities and screening on the stability of fractional quantum Hall states in graphene. 
We also show that the uppermost graphene layer can serve as a top gate, which can screen surface disorder and enable microscopy with displacement field control. This unlocks access to a wide range of phenomena that can only be observed in top-gated devices, from fractional Chern insulators in Moir\'e superlattices to correlated states in multilayer graphene.

\end{abstract}

\maketitle


\begin{figure*}[h!tbp]
    \includegraphics[width=0.9\textwidth]{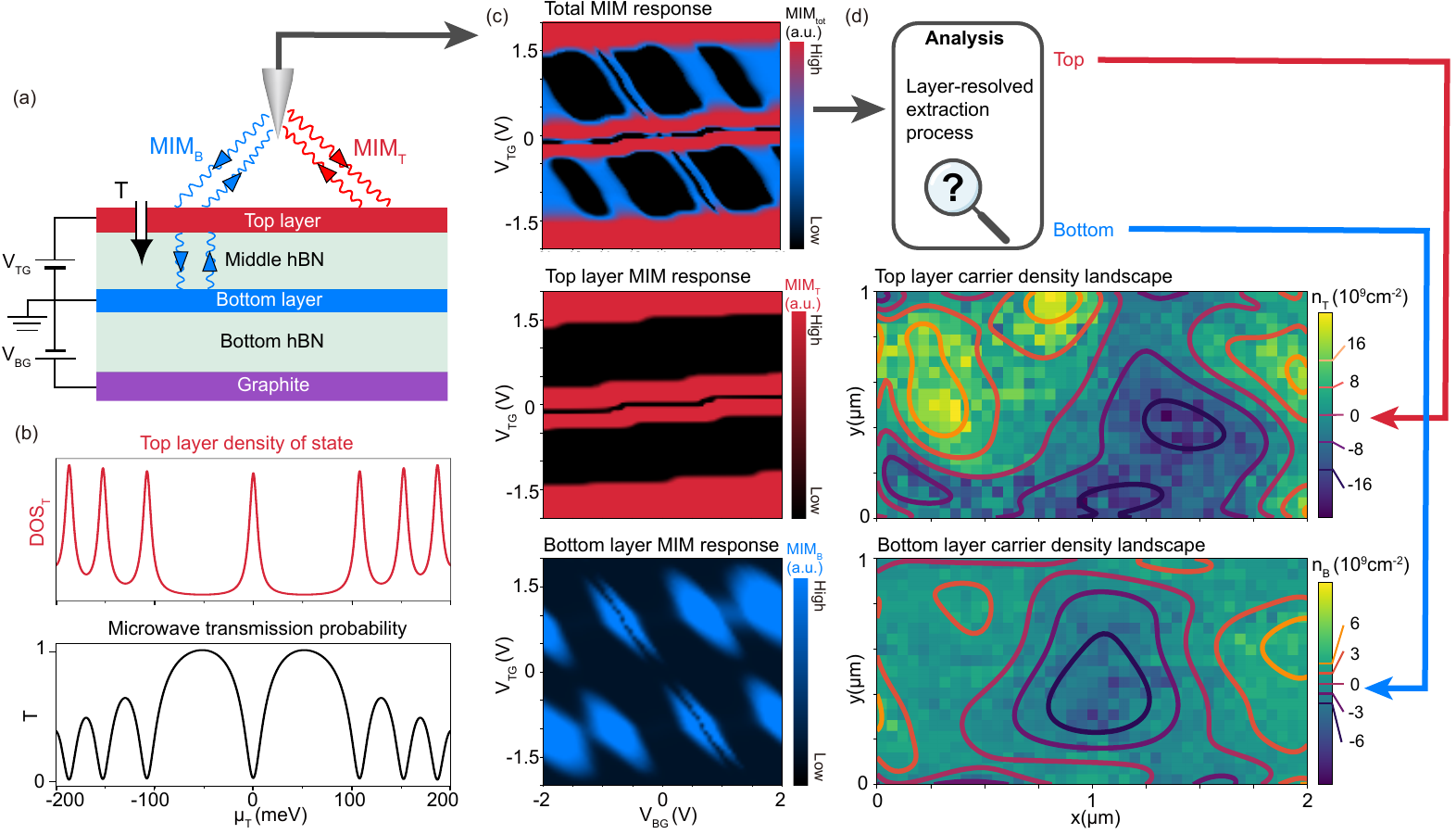}
    \caption{\label{fig:Motivation} 
    \textbf{Layer-resolved microwave impedance microscopy.} 
    \textbf{(a)} Schematic illustration of LR-MIM, where the measured signal ($\mathrm{MIM_{tot}}$) is a combination of the microwave response from the top layer ($\mathrm{MIM_T}$) and bottom layer ($\mathrm{MIM_B}$). In this work, the sample is a double-layer quantum Hall system.
    \textbf{(b)} Calculated density of states of the top layer ($\mathrm{DOS_{T}}$, red line) and the electromagnetic wave transmission probability ($\mathrm{T}$, black line)  as a function of the top layer chemical potential $\mu_\mathrm{T}$. The intensity of the interlayer microwave transmission exhibits a series of sharp minima associated with the quantized energy spectrum of the top layer.
    \textbf{(c)} \textit{Top panel:} Simulated total MIM signal to be detected by the tip.  \textit{Middle \& bottom panels:} Simulated contributions of $\mathrm{MIM_T}$ and $\mathrm{MIM_B}$ from the \textit{individual} layers as a function of the interlayer bias ($\mathrm{V_{TG}}$) and back gate voltage ($\mathrm{V_{BG}}$). 
    \textbf{(d)}    Disorder landscape of a double layer graphene device (a), imaged using layer-resolved microwave impedance microscopy (LR-MIM).
    3D imaging is challenging because states on all layers simultaneously contribute to the total microwave signal detected by the tip.
    This paper presents a process for layer-resolved extraction of the charge density, which can be used to map out the 3D disorder landscape and study its interplay with fractional quantum Hall phases on different atomic planes. 
    }
\end{figure*}

Surface-sensitive scanning probe microscopy (SPM) techniques are an exquisitely sensitive probe of strongly-correlated quantum matter. As a few recent examples, scanning tunneling microscopy (STM) studies have provided direct evidence of incommensurate density waves and superconducting gaps \cite{liu2024STM_DW,fan2025STM_SCgap}; single electron transistor (SET) probes of local compressibility have demonstrated symmetry-breaking including flavor polarization and fractional quantum Hall states \cite{feldman2012SETFQHE,feldman2013SETFlavor}; superconducting quantum interference device (SQUID) imaging has shown ``mosaics" of Chern insulators \cite{grover2022SQUIDMosaic}; and microwave-impedence microscopy (MIM) has directly measured conductive edge modes in topological insulators \cite{Lai20112DEG,Cui2016Gr,Shi2019WTe2,allen2019MIM}. 

However, many SPM techniques are restricted to imaging phenomena on the surface of a material \cite{martin1988EFM_surface,nonnenmacher1991KPFM_surface,yoo1997SSET_surface}, rendering states situated deep within the interior largely inaccessible. 
This presents challenges for the microscopy of van der Waals (vdW) heterostructures, as conductive states may simultaneously occupy multiple planes below the surface. Examples include: double-layer graphene or transition metal dichalcogenide stacks \cite{Zhang2009BLG,Cao2018MATBG, Wang2020WSe2WSe2,Li2021MoSe2WS2,Cai2023MoTe2,Park2024MoTe2,Ji2024MoTe2} (widely used for excitonic devices), moir\'e crystals with integrated sensor layers \cite{Xu2020WSe2WS2,Zeng2023MoTe2,Xia2024Sensor}, and dual-gated multilayer structures \cite{Enssline2018BGQD,Vandersypen2012BGhBN,Yacoby2012SBG,Nadj-Perge2023TLG,Ju2024CIPG,Ju2024FQAHMG,Young2024UCvdWLGG}.
Crucially, dual-gated geometries are required for many of the highest-quality devices, as this setup not only enables independent control of the carrier density and displacement field, allowing convenient tuning of correlated and topological phases, but also features significantly cleaner transport characteristics because the metallic top gate screens out surface disorder.
\cite{ong2012impurity,Zhang2009BLG,Cao2018MATBG,Wang2020WSe2WSe2,Li2021MoSe2WS2,Cai2023MoTe2,Park2024MoTe2,Ji2024MoTe2}. 
Imaging this rapidly expanding family of 
heterostructures presents a grand challenge, as the top layer typically blocks scanning probes from sensing the layer beneath it \cite{Li2021STM, Xie2021SET}.

This work introduces a 3D quantum imaging technique enabling \textit{layer-resolved} 
microwave impedance microscopy (LR-MIM)
 of van der Waals heterostructures in the mK regime. 
In addition to the technical challenges described above, development of this technique was motivated by a number of fundamental questions. For example: What is the vertical distribution of charges in a vdW heterostucture, and how is the surface disorder landscape imprinted onto different atomic planes below the surface? What is the interplay between the 3D disorder profile --- including out-of-plane impurities --- and fractional topological phases? Can we selectively manipulate the screening environment to tailor the disorder potential and interaction strength in a desired layer? We shall see that layer-resolved MIM is well-suited to address such questions.

Layer-resolved MIM works by separating the total MIM signal from all layers (Fig.~\ref{fig:Motivation}(a)) into layer-specific contributions (Fig.~\ref{fig:Motivation}(c)). The key idea is to open a ``microwave window" into the sample by tuning all layers except one into insulating gaps, so that MIM directly probes the selected layer (Fig.~\ref{fig:Motivation}(b)).
This technique naturally allows direct nanoscale imaging of 2D materials in top-gated geometries and is broadly applicable to multilayer electron systems, including double-layer platforms for excitonic superfluidity and beyond.

To demonstrate LR-MIM, we use a double layer graphene device in the quantum Hall regime as a testbed, separately resolving states on each atomic plane. 
By detecting shifts in the quantized energy levels of the top layer, we extract key features of the band structure and local thermodynamics in the bottom layer, including gap sizes and negative compressibilities, a hallmark of strong electron correlations. 
In the quantum Hall regime, the discrete energy spectrum of the top graphene layer
enables penetration of microwave photons through gapped states (Fig.~\ref{fig:Motivation}(b)), thus unlocking access to quantum phases in subsurface layers through a charged top gate. 
Using layer-resolved microwave readout, we image carrier density variations on each atomic plane and demonstrate that the top layer can be used to screen and attenuate potential fluctuations arising from surface disorder (Fig.~\ref{fig:Motivation}(d)). This screening behavior, governed by the local compressibility and interlayer coupling, plays a crucial role in stabilizing fragile fractional quantum Hall phases that are highly sensitive to the real-space disorder landscape. Furthermore, 3D disorder mapping helps to decouple intrinsic many-body effects from extrinsic influences such as strain, charge puddles and impurity-induced potentials.

LR-MIM operates in three complementary modes within the same device, with each sensitive to a different property of the sample.
\begin{itemize}
\item 
\textit{Mode 1: Direct surface microscopy}. This is ``standard" surface-sensitive MIM. In our testbed this reveals the strongly-correlated quantum Hall ferromagnets (QHFM) of the top layer (red in Fig.~\ref{fig:Motivation}(a)), as well as its real-space disorder profile, which is strong enough to wash out fractional states in this layer.
\item 
\textit{Mode 2: Indirect subsurface microscopy.} The top layer acts as a local quantum sensor for the inverse compressibility of the sample layer~\cite{eisenstein1992negative,Li2021STM_dual,Chiu2024STM_dual}. This information is imprinted onto the MIM signal of the top layer, which we use to extract quantitative gaps of the integer and fractional quantum Hall states. 
\item 
\textit{Mode 3: Direct subsurface microscopy.} Finally, we tune the top layer into a gap --- potentially with a finite carrier density providing displacement field --- and look through the microwave window to directly image the subsurface layer (blue in Fig.~\ref{fig:Motivation}(a)). We then directly measure the MIM response of the bottom layer itself, yielding an independent probe of the inverse compressibility, local admittance, and the local disorder profile. This provides a quantitative probe of the screening effect in dual-gates devices, which directly controls the stability of the fragile fractional states. 
\end{itemize}

The combination of these three modes builds a 3D map of the disorder profile, conductivity, and local thermodynamics of the sample. Because any 2D system can serve as the subsurface ``sample'' layer, these same modes of layer-resolved MIM can be extended to Moir\'e superlattices in twisted heterostructures, graphene multilayers, coupled quantum wells, and topological materials, offering a versatile approach for layer-resolved studies of correlated phases.
By integrating quantum sensing directly into the vdW device structure, this method bridges the gap between bulk measurements and surface probes, providing new opportunities to explore the spatial and electronic complexity of quantum materials in three dimensions.\\

\textbf{I. Experimental setup.}
We investigated a multilayer device consisting of a top layer (graphene), a bottom layer (graphene),
and a back gate (graphite), each separated by thin insulating spacers --- see Fig.~\ref{fig:Schematics}(a). 
Sample fabrication details are given in the Supplemental Information (SI). Leads adjust the potential differences $V_\text{BG}$ and $V_\text{TG}$ between the back gate or top layer and the bottom layer, respectively.

\begin{figure*}[h!tbp]
\includegraphics[width=0.9\textwidth]{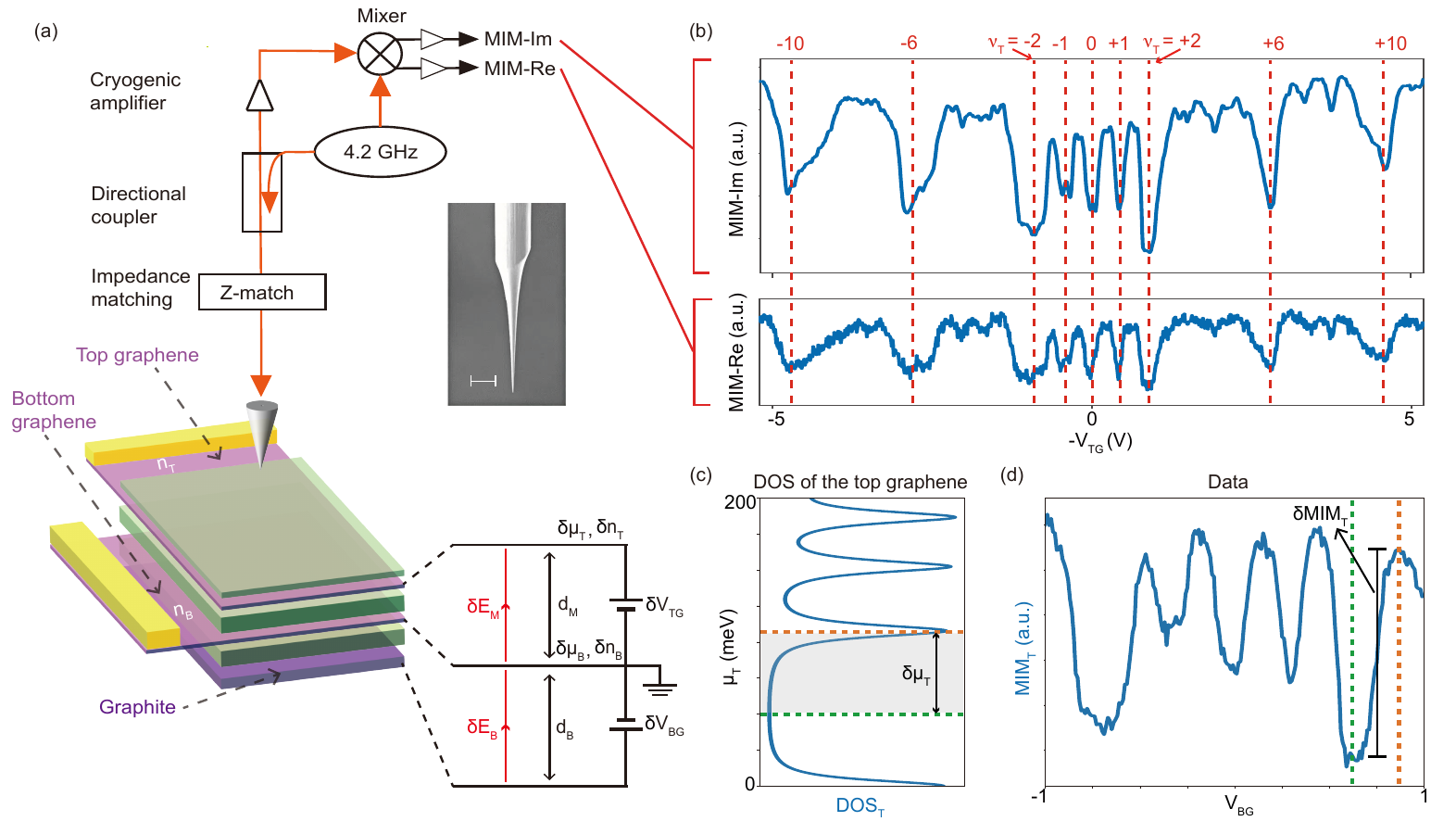}
\caption{\label{fig:Schematics} \textbf{Microwave impedance microscopy of double-layer graphene.} 
\textbf{(a)} 
The device consists of a top graphene layer, a bottom graphene layer, and a graphite back gate. 
Cross-sectional electrostatic profile of the conducting layers in the double-layer heterostructure is illustrated next to the device schematic.  
\textit{Inset:} Scanning electron micrograph of the MIM tip, prepared with electrochemical etching. Scale bar: \SI{20}{\mu m}. 
\textbf{(b)} Imaginary (MIM-Im, \textit{upper panel}) and real (MIM-Re, \textit{lower panel}) parts of the MIM response, measured as a function of inter-layer voltage bias, $V_\text{TG}$. $V_\text{BG} = 0V$. The tip is positioned over the bulk of the device, with a tip-sample distance of \SI{65}{nm}. 
\textbf{(c-d)} When a modulation $\delta \VBG$ is applied to the back gate, an electric field $\delta E_\mathrm{M}$ will penetrate the lower graphene layer (where the size of this penetration field depends on the density of states, DOS, of the lower layer). This shifts the chemical potential of the top layer, $\delta \mu_\mathrm{T}$ (panel c), which can be detected experimentally by measuring the change in the MIM response of the top layer, $\delta \mathrm{MIM}_\mathrm{T}$ (panel d). Panel d shows the back-gate dependence when the lower graphene layer is kept at low DOS.
}
\end{figure*}

The double-layer graphene sample was characterized using microwave impedance microscopy (MIM) at temperatures down to \SI{60}{mK}
in a dry dilution fridge with a superconducting magnet,  shown in Fig.~\ref{fig:Schematics}(a) \cite{Cao2023mK}. 
Our imaging system consists of a metallic scanning probe that probes the charge susceptibility $\chi$ (i.e. the dynamical density-density correlator) of the sample using RF reflectometry. 
When the tip is suspended a distance $d$ above the sample, the imaginary part of the MIM response is given in linear response by~\cite{Taige2023comsol}, 
\begin{equation}
    \Im \mathrm{MIM} 
    \propto \int_{\bq} \chi(\omega\approx 0, \bq) \frac{e^{-2|\bq| d}}{|\bq|^2},
    \label{eq:MIM_linear_response}
\end{equation}
where $\chi(\omega, \bq)$ is the real part of the charge susceptibility and $\bq$ is the momentum in the sample plane. 
Here $\omega = 4.22$ \si{GHz}. 
Physically, this response comes from microwaves scattering into the sample and creating a charge density fluctuation. Those fluctuations create outgoing near-field microwaves, which scatter back to the tip and are measured.
This implies that for typical tip-sample geometry, at the leading order, the response is proportional to the charge compressibility of the material~\cite{Taige2023comsol}.

We will now discuss the MIM response of the full multilayer system. 
Crucially, the microwaves used in MIM partially penetrate through the top layer, producing a reflected signal that includes information about \textit{both} layers (Fig.~\ref{fig:Motivation}(a)).
In particular, the MIM response of the multilayer is the sum of the top layer response (given by Eq.~\eqref{eq:MIM_linear_response} in SI)
plus the response of the bottom layer: $\mathrm{MIM}_{\mathrm{tot}} = \mathrm{MIM}_{\mathrm{T}} + \mathrm{MIM}_{\mathrm{B}}$. 
Physically, the second term comes from microwaves scattering from the tip then \textit{through} the top layer and the hexagonal boron nitride (hBN) spacer to create a charge fluctuation in the bottom layer. This creates outgoing microwaves that make the reverse journey. We model this schematically as 
\begin{equation}
    \mathrm{MIM}_{B} \propto \int_{\br,\br'} \mathcal{G}_{\br_t,\br} \chi_B^{\omega,\br,\br'} \mathcal{G}_{\br',\br_t}; \hspace{0.5em} 
    \mathcal{G} =  G^{\mathrm{vac}} \mathsf{T}[\sigma_{\mathrm{T}}] G^{\mathrm{hBN}},
    \label{eq:subsurface_MIM_response}
\end{equation}
where $\chi_B$ is the charge susceptibility of the lower layer, $\br_t$ is the tip location, $\br,\br'$ are positions within the plane of the sample,  and $\mathcal{G}$ is the classical electromagnetic Green's function for microwave photons to go from the MIM tip to the lower layer. The Green's function is broken into three parts: (1) scattering from the tip to the top layer through vacuum, $G^{\mathrm{vac}}$, (2) optical transfer through the top layer, $\mathsf{T}$, and (3) scattering from the top layer through the hBN to the bottom layer, $G^{\mathrm{hBN}}$. The optical transfer matrix $\mathsf{T}[\sigma_{\mathrm{T}}]$ is a functional of the conductivity tensor of the top layer $\sigma_\mathrm{T}$, which determines the extent to which it screens the microwaves from passing through. When the top layer is in an incompressible state whose gap is above the \si{\giga\hertz} scale, it becomes transparent: $\mathsf{T} \to \textrm{Id}$, as shown in Fig.~\ref{fig:Motivation}(b). Tuning the Fermi energy of the top layer into an insulating gap therefore opens a ``window" through which the MIM signal from the subsurface layer can be directly detected.

To extract the wealth of information contained in this seemingly-complicated signal, we will investigate the contributions to the overall response from each layer separately. For clarity, we first discuss the signal from the top layer, $\mathrm{MIM}_{\mathrm{T}}$, which dominates the total MIM response. The $\mathrm{MIM}_{\mathrm{T}}$ signal already contains \textit{indirect} information about the bottom layer, namely its inverse compressibility. The subsequent section will focus on extracting and isolating the \textit{direct} contribution from the subsurface layer, $\mathrm{MIM}_{\mathrm{B}}$.
To understand these experimental results, we will verify the linear response model above to high accuracy with finite element simulations, allowing us to understand the origin of each part of the signal. With this interpretation in place, it will become clear 
how to perform layer-resolved imaging and 
to \textit{directly} probe properties of the subsurface sample layer.\\

\textbf{II. Experimental results.}
We locally probed the MIM response of the double-layer system with the tip positioned away from the edge to measure the bulk response as a function of  the back gate voltage $\VBG$ and the inter-layer bias $\VTG$.

Initially we set $\VBG=0$ so that $n_\text{T} \propto V_\text{TG}$ approximately and the MIM response is dominated by the top layer (subsurface MIM response is discussed below). Fig.~\ref{fig:Schematics}(b) shows the measured imaginary (MIM-Im)  and real (MIM-Re) parts of the microwave response as a function of $V_\text{TG}$ in a magnetic field of \SI{8}{T}. Dips in the response indicate incompressible states where $dn/d\mu$ vanishes. The largest dips at fillings $\nu = \pm 2, \pm 6, \pm 10,\dots$ correspond to filling the four-fold degenerate Landau levels of monolayer graphene~\cite{castro2009electronic,Cui2016Gr}. 
Smaller dips at $\nu = -1,0,1$ within the zeroth Landau level are the quantum hall ferromagnets (QHFM) of graphene resulting from strong interactions leading to spontaneous breaking of spin and valley symmetry~\cite{Halperin2013} .
QHFM states are also visible in some of the higher Landau levels. Only subtle signatures of fractional states are visible, which we attribute to surface disorder on this non-top-gated monolayer. Altogether, these measurements confirm the MIM response of the top layer is exactly as expected from monolayer graphene.

\begin{figure*}[h!tbp]
\centering
\includegraphics[width=0.9\textwidth]{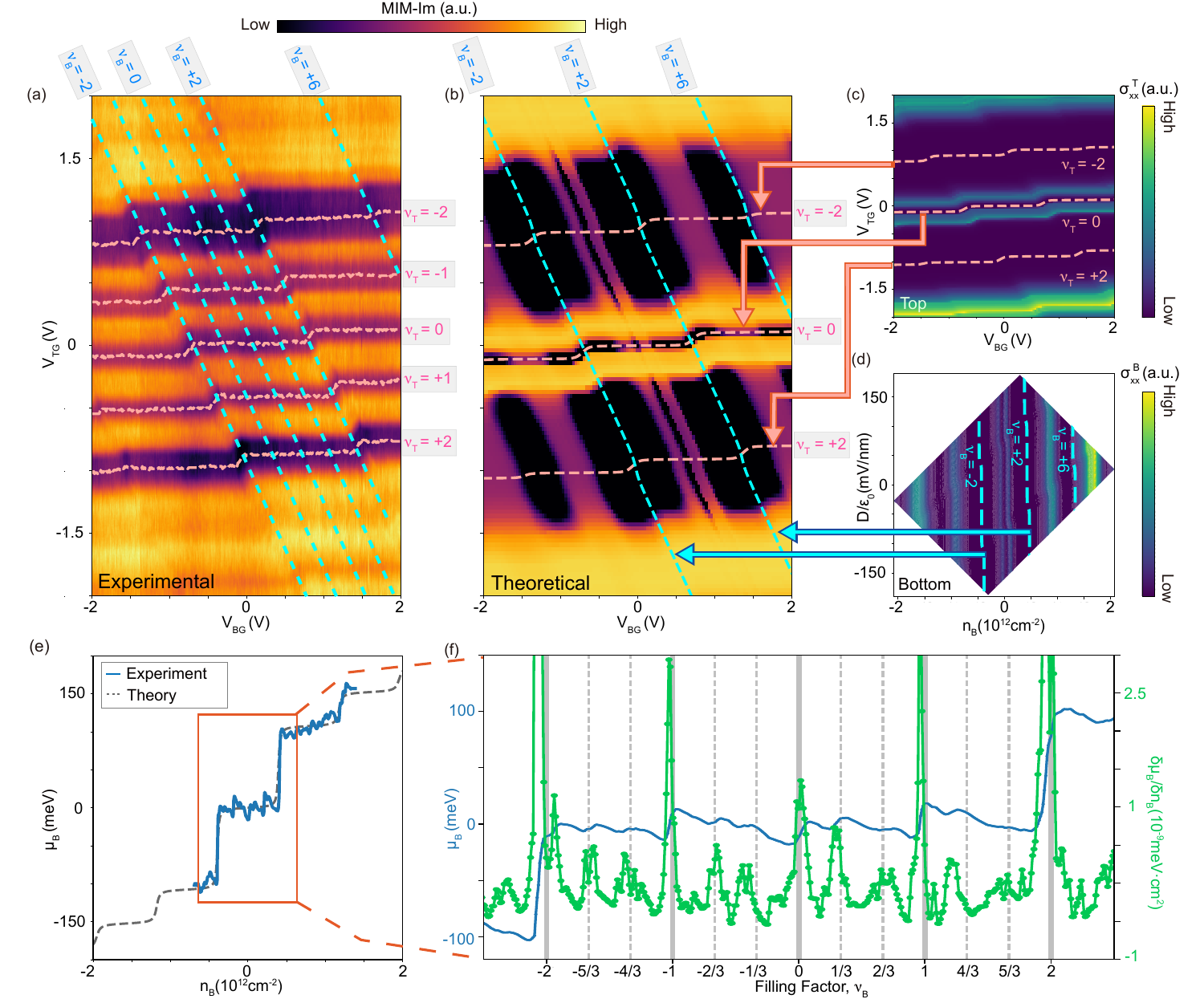}
\caption{\label{fig:LL}  
\textbf{Multilayer MIM: local compressibility measurement of a fractional quantum Hall insulator with displacement field control.}
\textbf{(a-b)} Experimental (\textit{left}) and theoretical (\textit{right}) plots of the MIM-Im response of the double-layer system, plotted as a function of $V_\text{TG}$ and $V_\text{BG}$. The experimental data was obtained at at $T$ = \SI{60}{mK} and $B$ = \SI{8}{T}. The corresponding filling factors $\nu_{\text{T}}$ (top layer) and $\nu_{\text{B}}$ (bottom layer) are labeled in pink and blue, respectively. 
\textbf{(c-d)} Simulated longitudinal conductivities of the top ( $\sigma^{xx}_\text{T}$, \textit{panel c}) and bottom ($\sigma^{xx}_\text{B}$, \textit{panel d}) graphene layers. The correspondence between the filling factors observed in the MIM-Im simulation in (b) and the longitudinal conductivities in (c-d) is indicated by arrows. The axes in (d) correspond to the vertical displacement field ($D/\epsilon_0$) and the bottom layer carrier density ($n_\mathrm{B}$).
\textbf{(e)} Chemical potential of the bottom graphene layer, $\mu_\text{B}(n_{\text{B}})$, extracted experimentally from the $\nu_{\text{T}} = -2 $ trajectory in (a), plotted in blue solid lines. 
The chemical potential predicted by theory is plotted with a gray dashed line.
\textbf{(f)} Chemical potential (\textit{left}, blue) and inverse compressibility (\textit{right}, green, $\delta\mu_\text{B}/\delta n_\text{B}$) plotted as a function of the bottom layer filling factor, $\nu_\text{B}$. 
}
\end{figure*}



Fig. \ref{fig:LL}(a) shows the measured imaginary MIM signal as a function of $\VTG$ and $\VBG$. We will first describe the effect of the back gate and inter-layer bias voltages on the carrier densities of both layers and then show how the MIM signal manifests. 
The top layer acts as an unusual top gate. Unlike a conventional metallic or graphite top gate, its density of states is comparable to that of the lower ``sample'' layer. Changing the carrier density $n_\text{B}$ of the bottom layer therefore induces a change in the carrier density $n_\text{T}$ of the top layer. To understand this behavior, we must consider how the density of states within each layer impacts the out-of-plane electric field profile, and vice versa \cite{Eisenstein1994compressibility}.
If we start with both the top and bottom layers at the charge neutrality point (CNP), a voltage perturbation $\delta V_{\text{BG}}$ or $\delta V_{\text{TG}}$ will induce a perpendicular displacement field and shift the chemical potential in both layers:
$e \delta \VTG = -e d_{\text{M}} \delta E_{\text{M}} + (\delta \mu_{\text{B}} - \delta \mu_{\text{T}})$ and
$e \delta \VBG = e d_{\text{B}}  \delta E_{\text{B}} + \delta \mu_{\text{B}}$. 
Here $\delta \mu_{\text{B}}$ and $\delta \mu_{\text{T}}$ are the chemical potential changes in the bottom and top layers, and $d_{\text{B}}$ = 32 nm and $d_{\text{M}}$ = 42 nm are the thicknesses of the bottom and middle hBN layers, respectively (Fig~\ref{fig:Schematics}(a)). $\delta E_{\text{B}}$ and $\delta E_{\text{M}}$ are the changes in the perpendicular displacement fields in the corresponding hBN dielectric layers. These are related to the charge densities by $\delta E_{\text{B}} = (e/\epsilon) (\delta n_{\text{T}} + \delta n_{\text{B}})$ and $\delta E_{\text{M}} = (e/\epsilon) \delta n_{\text{T}}$ with electric permittivity $\epsilon$, as illustrated in the electrostatic diagram
in Fig.~\ref{fig:Schematics}(a). 
Therefore, the relations between the applied gate voltages and the carrier densities in the two layers can be expressed as \cite{Tutuc2012Fermi}:
\begin{subequations}
\label{eq:voltage_carrier_relations}
\begin{align}
\label{eqn:Tutuc1}
e\VTG &= [\mu_\text{B}(n_{\text{B}}) - \mu_\text{T}(n_{\text{T}}) ] - e^2 n_{\text{T}}/ C_{\text{M}},\\
\label{eqn:Tutuc2}
e\VBG &= \mu_\text{B}(n_{\text{B}}) + e^2(n_{\text{B}}+n_{\text{T}})/C_{\text{B}}
\end{align}
\end{subequations}
where $C_{\text{M}} = \SI{72}{\nano\farad\per\centi\meter\squared}$ and
$C_{\text{B}} = \SI{95}{\nano\farad\per\centi\meter\squared}$ are the dielectric capacitances per unit area of the middle and bottom hBN, respectively. 
In this case, the top layer is monolayer graphene, which has a known chemical potential $\mu_\text{T}(n_\text{T})$. At the single particle level, the chemical potential of graphene in a magnetic field $B$ is given by
$\mu(n) = \pm v_F \sqrt{2 \hbar e B \left| N(n) \right| }$ where $N(n)$ is the number of filled Landau levels. Equipped with the relationship between voltage and carrier density, we now move to microscopy.

Fig. \ref{fig:LL}(a) shows the measured imaginary MIM signal (MIM-Im) as a function of $\VTG$ and $\VBG$. As in Fig.~\ref{fig:Schematics}(b), tuning $\VTG$ mostly adjusts $n_\text{T}$, producing minima in the MIM-Im signal whenever the top layer becomes incompressible. The trajectories of these minima are marked by the pink dotted lines, and correspond to \textit{top layer} fillings $\nu_\text{T} = -2,-1,0,1,2$~\cite{kim2021AFM}. As $\VBG$ increases, these Landau level trajectories move to higher $\VTG$ following a staircase pattern, with regions of shallow slope interrupted by discrete jumps at specific gate voltages. These jumps can be understood as follows: when the sample layer is in a gapped state, the electric field from a back gate modulation $\delta \VBG$ will penetrate through the sample and shift the chemical potential  $\delta \mu_\text{T}$ of the upper layer, creating in a sharp jump in the Landau level trajectory of the upper layer, as illustrated in Fig.~\ref{fig:Schematics}(c-d). When the sample is in a compressible state, by contrast, it will (partially) screen the field from the back gate, resulting in a smooth trajectory. Therefore the jumps in incompressible states of the top layer occur exactly at the incompressible states of the sample layer. We may identify these as the states $\nu_\text{B}=0, \pm 1, \pm 2, \pm 6, \dots$ of the sample layer (blue lines). The pattern of minima in Fig. \ref{fig:LL}(a) is therefore governed by the relationship between voltage, chemical potential, and carrier density in Eq.~\eqref{eq:voltage_carrier_relations}.

To understand and validate these results from a theoretical perspective, we turn to simulations of the linear response model of multilayer MIM introduced in Section I. The only quantum part of 
Eqs.~\eqref{eq:MIM_linear_response} \& ~\eqref{eq:subsurface_MIM_response} (see SI) are the charge susceptibilities or, equivalently, the conductivities of the top sensor layer ($\sigma_{\mathrm{T}}$) and the bottom sample layer ($\sigma_{\mathrm{B}}$). For simplicity, these conductivities are calculated from the transport data presented in the reference \cite{novoselov2005two}, including disorder-induced broadening of the Landau levels (see SI for a detailed discussion).
This fully determines the charge densities $n_{\mathrm{\ell}}(\mu_{\ell})$ and conductivities $\sigma_{\ell}(n_\ell)$ for $\ell = \mathrm{T}, \mathrm{B}$. 
We then use Eqs.~\eqref{eq:voltage_carrier_relations} to determine the carrier densities $n_{\mathrm{T}}(\VBG,\VTG)$ and $n_{\mathrm{B}}(\VBG,\VTG)$. 
The resulting longitudinal conductivities $\sigma^{xx}_{\mathrm{T}}(\VBG,\VTG)$ and $\sigma^{xx}_{\mathrm{B}}(\VBG,\VTG)$ are shown in Fig. \ref{fig:LL}(c-d), where pink and blue trajectories indicate the centers of the incompressible gaps in the top and bottom layers, respectively. These correspond closely to the observed experimental trajectories in Fig.~\ref{fig:LL}(a) --- with the key difference that quantum Hall ferromagnetism is not present at the single particle level, so the $\nu= \pm 1, \pm 3,\dots$ states are not included. 
Finally, the theoretical multilayer MIM response can be computed from these longitudinal conductivities, using classical electromagnetic simulations incorporating the device geometry and dielectric environment \cite{Taige2023comsol} (See SI).
A side-by-side comparison of the experimental (Fig.~\ref{fig:LL}(a)) and simulated (Fig.~\ref{fig:LL}(b)) MIM signals shows a close correspondence. This constitutes strong evidence that the multilayer MIM signal is described by the combination of the voltage-carrier relation and the linear response theory of MIM.
 
\textit{\textbf{Indirect Subsurface Microscopy.}}
The Landau level trajectories of the top ``sensor" layer can be used to measure the inverse compressibility of the subsurface ``sample'' layer, thereby enabling quantitative readout of energy gaps in the sample. Remarkably, when the top graphene layer is at filling $\nu_{\text{T}} = 0$, Eq.~\eqref{eq:voltage_carrier_relations} simplifies to $eV_{\text{TG}} = \mu_{\text{B}}(n_{\text{B}})$. Thus, by tracking the voltage coordinates $(V_\text{BG}^i, V_\text{TG}^i)$ that trace the Landau level trajectories of the top layer (dotted pink lines in Fig.~\ref{fig:LL}(a)) and plugging them into Eqs.~\eqref{eq:voltage_carrier_relations}
one can immediately extract the Fermi energy of the bottom layer. 
(See the SI for a discussion of the generalization of this concept to higher Landau levels and the accompanying data analysis.)
Following this procedure, one can locally extract $\mu_\text{B}(n_\text{B})$ for the bottom graphene layer, as shown in Fig.~\ref{fig:LL}(e). We resolve large jumps in $\mu_\text{B}$ at carrier densities corresponding to $\nu_\text{B}=\pm2$. This indicates a gap size of $\Delta \sim \SI{105}{meV}$, which is consistent with energy scales expected theoretically and validates the picture above (quantified in Eqs.~\eqref{eq:voltage_carrier_relations}).
Upon closer inspection of the experimental $\mu_\text{B}(n_\text{B})$ data, clear deviations from single-particle theory are apparent at low carrier densities (Fig.~\ref{fig:LL}(e), orange box). Zooming into this region and plotting the inverse compressibility $\kappa_\text{B} = \delta\mu_\text{B}/\delta n_\text{B} $ (Fig.~\ref{fig:LL}(f)), one can resolve signatures of broken symmetry quantum Hall states ($\nu_{\text{B}}=0, \pm1$) and hints of fractional fillings, such as $\nu_{\text{B}} = \pm 1/3$ and $\pm 2/3$. A negative slope appears around the quantum Hall ferromagnets $\nu_{\text{B}}=0, \pm1$, corresponding to a region of negative compressibility driven by interactions. This is consistent with kPFM, SET, STM, and other findings \cite{eisenstein1992negative,park2020flavour, zondiner2020cascade, kim2021AFM}. 

\begin{figure*}[ht!]
\includegraphics[width=0.9\textwidth]{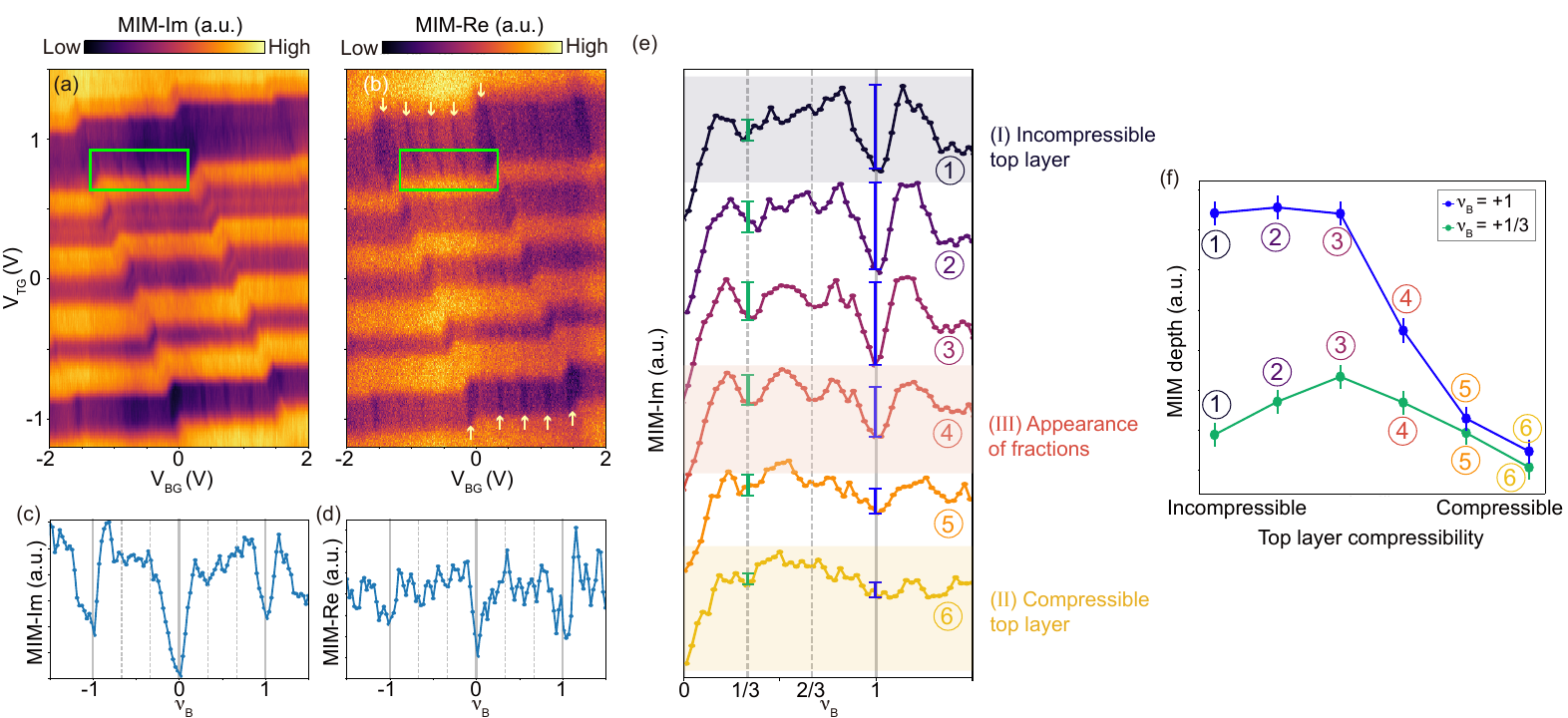}
\caption{\label{fig:Comparison} 
\textbf{Subsurface MIM of quantum Hall states in graphene and stabilization of the FQHE by disorder screening.} 
\textbf{(a-b)} Measured MIM-Im (a) and MIM-Re (b) as a function of $V_\text{TG}$ and $V_\text{BG}$ in the quantum Hall regime. Measurements were performed at $T$= \SI{60}{mK} and $B$ = \SI{8}{T}. Yellow arrows in (b) indicate the $\nu_{\text{B}} = 0,\pm1,\pm2$ bottom layer filling factors. 
\textbf{(c-d)} Line cut of the MIM-Im signal (c) and MIM-Re signal (d) extracted 
when top graphene layer is in the  $\nu_{\text{T}} =  +2$ incompressible state. In (c-d), the broken symmetry states of the bottom graphene layer, $\nu_{\text{B}} = 0,\pm1$, are fully developed. Fillings where the fractional states $\nu_{\text{B}} = \pm 1/3, \pm 2/3$ would emerge are marked by gray dotted lines.
\textbf{(e)} Evolution of the integer and fractional fillings in the lower layer as the compressibility of the top layer is varied. The series of line cuts are extracted around the transition of the $\nu_{\text{T}} = -2$ state in the top layer, shown with the green box in (a-b). Fractional fillings  $\nu_{\text{B}} = \pm 1/3, \pm2/3$ are marked with gray dotted lines.
\textbf{(f)} Evolution of the strength of the integer $\nu_\mathrm{B}=1$ (\textit{blue line}, defined as the amplitude of the incompressible $\nu_\mathrm{B}=1$ dip in panel (e) and fractional $\nu_\mathrm{B}=1/3$ (\textit{green line}) states in the bottom layer, plotted as a function of the compressibility of the top layer. The FQHE in the lower layer appears to become stabilized as the conductivity of the top layer increases (data points 2-3), likely due to screening of surface disorder. When the top layer is highly compressible (data points 5-6), it begins to screen the microwave signal from the tip, leading to a suppression of the MIM response of both integer and fractional states.}

\end{figure*}


\textbf{\textit{Direct Subsurface Microscopy.}}
In this experiment, the microwave signatures of electronic states on the bottom graphene layer are subtle and challenging to extract directly from the data, as the overall signal is dominated by the layer closest to the tip. 
Using the longitudinal conductivities in Fig.~\ref{fig:LL}(c-d) as a reference, we can identify the features in the overall MIM response (Fig.~\ref{fig:LL}(a-b)) that are  expected to arise from Landau levels in the top layer (labeled in pink) or the bottom layer (labeled in blue).
As shown in the plot of $\sigma^{xx}_\text{B}(\VBG,\VTG)$ in Fig.~\ref{fig:LL}(d), the  lower layer can be treated roughly like a dual-gated sample in which the upper layer functions as a ``top gate''. 


\begin{figure*}[ht!]
\includegraphics[width=0.9\textwidth]{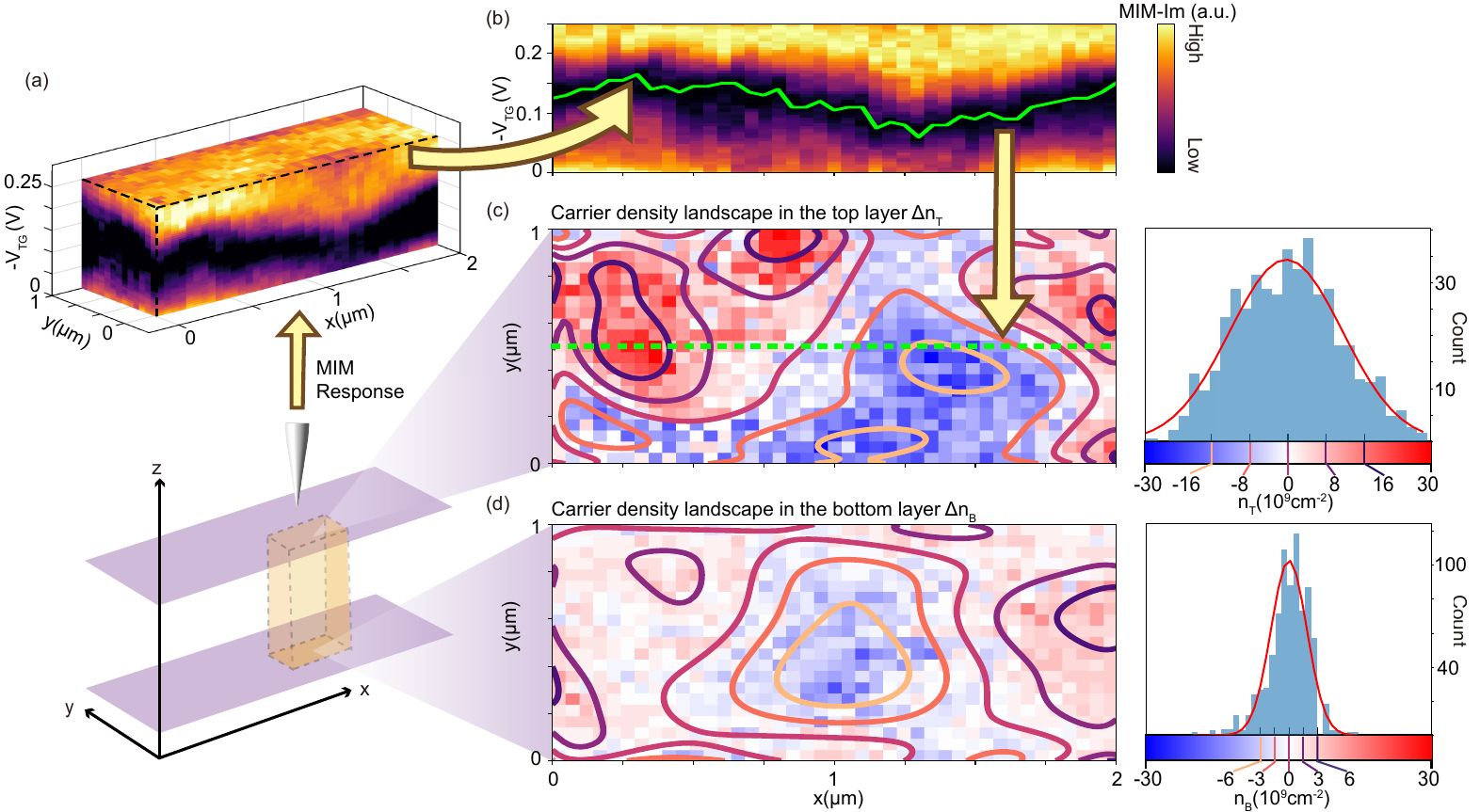}
\caption{\textbf{Layer-resolved imaging of the charge density in a vdW heterostructure.}
\textbf{(a)} Schematic illustration of the double-layer sample and MIM readout in the x-y-$V_\text{TG}$ coordinates.  
\textbf{(b)} Measured MIM-Im response (a.u.) as a function of interlayer bias $V_{\text{TG}} $ and position, which tracks small shifts of the $\nu_{\text{T}} = 0$ dip as the tip is scanned across the x-axis of the sample (green line).
\textbf{(c-d)} Spatial carrier density profiles for the top ($\Delta n_\mathrm{T}$, 
panel (c)) and bottom ($\Delta n_\mathrm{B}$, panel (d)) graphene layers. Histograms of the $n_\text{T}$ and $n_\text{B}$ density distributions are on the right. Panels (b-d) share the same x-axis, and contour lines correspond to the carrier densities labeled in the histograms. The red curves on the histograms are Gaussian fits. 
} 
\label{fig:Local}
\end{figure*}

Guided by these insights, we isolated the MIM response of the electronic states in the lower layer by imaging \textit{through} the upper layer when it is tuned to an incompressible state. The concept is illustrated in the theoretical MIM plot in Fig.~\ref{fig:LL}(b), which shows that the visibility of Landau level oscillations in the lower layer should be maximized when the top layer is fully gapped (horizontal dark stripes).
Similarly, our experimental MIM results reveal that the quantum Hall states in the bottom graphene flake manifest themselves as the black vertical stripes superimposed on the horizontal purple bands (Fig.~\ref{fig:Comparison}(a-b)). Direct MIM readout of electronic states on the \textit{bottom layer} are clearly visible as dips at $\nu_{\text{B}} = -1,0,1$ in one-dimensional line cuts of the MIM-Im and MIM-Re plots Fig.~\ref{fig:Comparison}(c-d).\\

\textbf{III. 3D microwave imaging of quantum states in a two-layer system.}
In this final section, we combine the developments above to perform \textit{layer-resolved microscopy}. 
Although MIM is sensitive to order, as evidenced by the FQHE states studied above, we now turn our focus to \textit{disorder}, creating a map of charge disorder profile on both the surface and subsurface levels. We will see that this novel technique immediately sheds light on longstanding experimental and theoretical questions about the stability of fragile many-body states, including their interplay with the disorder landscape.

We performed layer-resolved microscopy by combining the subsurface imaging demonstrated above with the lateral motion of the tip. To recap, we showed that, at a fixed location $(x,y)$, the MIM signal allows us to determine, \textit{inter alia}, the carrier densities $n_\mathrm{T}$ and $n_\mathrm{B}$ where many-body gaps occur on each layer. Due to charge disorder, we expect both the many-body gaps and the midgap carrier densities to become spatially-disordered. 
We mapped this out quantitatively starting in Fig.~\ref{fig:Local}(b). There we measured the MIM signal across the $\nu_\text{T} = 0$ plateau transition as a function of $\VTG$ along a \SI{2}{\micro\meter} linear path, showing the spatial variation of the midgap voltage $\Delta \VTG(x)$ (green line). The critical density $\Delta n_\mathrm{T}$ is related to 
$\Delta \VTG$ by $e\Delta V_{\text{TG}}(x) = - e^2 \Delta n_{\text{T}}(x)/ C_{\text{M}}$, furnishing quantitative maps of its spatial profile. 
By expanding the spatial scan into a two-dimensional grid, we can map out the full carrier density profile on the top layer (Fig.~\ref{fig:Local}(c)). Using the method to isolate the MIM response of Landau levels in the bottom layer (Fig.~\ref{fig:Comparison}(c)), we also constructed a real space map of the critical density $\Delta n_{\text{B}}(x,y)$ on the lower layer, shown in Fig.~\ref{fig:Local}(d).

If we compare the individual density distributions on the two layers, it appears that the geometry of the disorder potential on the surface layer is imprinted onto the buried layer, but with a strongly attenuated amplitude.
To quantify this, histograms of the upper and lower density distributions accompany each image in Fig.~\ref{fig:Local}(c-d): the standard deviations are around $0.0118\times 10^{-12} \,{\rm cm}^{-2} $ and $0.0038\times 10^{-12}\,{\rm cm}^{-2}$, respectively.
This suppression of charge disorder below the surface could be attributed to a combination of factors, including electrostatic screening of surface disorder by the upper graphene layer and the depth of the lower layer below the surface. Separately, layer-resolved MIM measurements have shown that FQH states are more stable on the bottom than the top layer, proportional to disorder. 
This is likely due to the fact that the top graphene layer can screen away the random disorder potential on the surface, introduced by nanofabrication polymer residues, surface impurities, etc.

\textbf{IV. Stabilization of the FQHE by Disorder Screening.}
To investigate how screening of surface disorder may impact the formation and visibility of integer and fractional quantum Hall states, we probed the lower layer while tuning the conductivity of the top layer. The results are displayed in Fig.~\ref{fig:Comparison}(e). The general trend is that the integer fillings, $\nu_{\text{B}} = -1,0,1$, are clearly visible in the MIM-Im response when the top graphene layer is insulating and become harder to resolve as it becomes more conductive. We observe three different regimes of behavior: 
(I) When the top layer is fully gapped (i.e., no screening of surface disorder), we can easily image through the top layer and observe prominent dips at fillings $\nu_\text{B} = -1,0,1$ with excellent visibility.
(II) When the top layer is fully conductive, a small dip in the MIM-Im channel can be resolved at $\nu_\text{B}=0$, but the visibility of all other integer and fractional states is suppressed. This likely occurs because the highly compressible top layer screens the fields from the tip, reducing the contributions of the bottom layer to the reflected RF signal.
(III) In an intermediate regime where the top layer is partially compressible, we observe fractional fillings $\nu_\text{B}=\pm 2/3, \pm 1/3$ in the lower layer only. This likely arises because the moderately conductive top layer screens surface disorder but is just insulating enough to allow the MIM signal to penetrate through it.

The disorder landscape governing the MIM data here is closely connected to an old but unresolved problem: plateau transitions. In the presence of disorder, IQH and FQH states persist over a range of densities or, equivalently, magnetic fields, creating a ``plateau" in the resistivity $R_{xy}$. At the edge of the plateau, the resistivity has a singular behavior $(dR_{xy}/dB)_{\mathrm{max}} \sim T^{-\nu z}$, where $z=1$ is the dynamical scaling exponent and experiments have measured $\nu \approx 2.38$~\cite{PhysRevLett.102.216801}. Theoretically, this has been understood as a consequence of disorder creating a random potential landscape, with electrons constrained to move close to equipotential lines~\cite{arovasFQHEnotes}. The plateau transition then occurs when the random landscape fills up enough to allow (quantum) percolation across the sample. A long line of theoretical work (reviewed in~\cite{PhysRevB.99.121301}) has studied this transition since the Chalker-Coddington model in the late 1980s~\cite{chalker1988percolation}, arriving at critical exponents around $\nu \approx 2.58-2.59$ --- close but ultimately distinct from the experimental value \cite{PhysRevLett.102.216801}. So, surprisingly, plateau transitions are poorly understood after four decades of study.

The subsurface MIM in~Fig.~\ref{fig:Local} provides direct evidence for the random disorder potential assumed in theories of plateau transitions. Surprisingly, despite the real-space character of plateau transitions, extremely limited microscopy data appears to be available. To our knowledge, previous microscopy studies on the LL regime of graphene, such as those with STM~\cite{li2013evolution,coissard2023absence,liu2022visualizing} or KPFM ~\cite{kim2021AFM} have focused on the edge or defects rather than the bulk. Subsurface MIM is sensitive to the potential landscape itself, and we showed above that it exhibits Gaussian potential fluctuations around a mean --- a basic input to any theory of the plateau. Future work will examine this problem more quantitatively, making direct measurements of the real-space fluctuations to potentially provide an independent measurement of the critical exponents. As MIM is sensitive to the local conductivity, it may even be possible to see signatures of the network of percolating conductive channels near criticality. Subsurface MIM should therefore provide a new window into the old-but-unresolved problem of the plateau transitions.

\textbf{Outlook.} The subsurface capabilities of layer-resolved MIM open up new opportunities to probe correlated phases in vdW heterostructures.   
One application is to
dual-gated devices, where subsurface imaging permits microscopy through the top gate to the sample below. This also enables microscopy in a separate class of topological states in epitaxial semiconductors that cannot easily be back gated, such as cadmium arsenide \cite{rashidi2023Cd3As2,munyan2023Cd3As2} and magnetically doped (Bi,Sb)Te \cite{mogi2017BiSbTe,allen2019MIM}.
While our approach in this work used the convenient Landau level gaps of monolayer graphene to provide displacement field and modulate microwave transmission, magnetic fields are not required. Indeed LR-MIM could be performed on layered structures with intrinsic bandgaps at zero magnetic field, opening the door to studying excitonic states, quantum spin Hall insulators, Wigner crystals, and beyond. Finally, the disorder mapping capabilities demonstrated here support a wide range of device characterization applications, ranging from investigations of the interplay between disorder and delicate strongly-correlated quantum states to tuning the screening environment of mesoscale quantum electronics. Layer‐resolved MIM can thus become a powerful tool for exploring emergent quantum phases and guiding future device technologies.





\hfill
  
\begin{acknowledgments}

We thank Tarun Grover and Yi-Zhuang You for inspiring discussions.
We gratefully acknowledge funding support from the UC Office of the President, specifically the UC Laboratory Fees Research Program (award LFR-20-653926), the AFOSR Young Investigator Program (award FA9550-20-1-0035), the AFOSR/ARO MURI Program (award FA9550-22-1-0270), and the NSF CAREER program (award DMR-2240114). This work was performed, in part, at the San Diego Nanotechnology Infrastructure (SDNI) of UCSD, a member of the National Nanotechnology Coordinated Infrastructure, which is supported by the National Science Foundation (Grant ECCS-2025752).
\end{acknowledgments}

\bibliography{main.bib}

\begin{thebibliography}{58}%
\makeatletter
\providecommand \@ifxundefined [1]{%
 \@ifx{#1\undefined}
}%
\providecommand \@ifnum [1]{%
 \ifnum #1\expandafter \@firstoftwo
 \else \expandafter \@secondoftwo
 \fi
}%
\providecommand \@ifx [1]{%
 \ifx #1\expandafter \@firstoftwo
 \else \expandafter \@secondoftwo
 \fi
}%
\providecommand \natexlab [1]{#1}%
\providecommand \enquote  [1]{``#1''}%
\providecommand \bibnamefont  [1]{#1}%
\providecommand \bibfnamefont [1]{#1}%
\providecommand \citenamefont [1]{#1}%
\providecommand \href@noop [0]{\@secondoftwo}%
\providecommand \href [0]{\begingroup \@sanitize@url \@href}%
\providecommand \@href[1]{\@@startlink{#1}\@@href}%
\providecommand \@@href[1]{\endgroup#1\@@endlink}%
\providecommand \@sanitize@url [0]{\catcode `\\12\catcode `\$12\catcode `\&12\catcode `\#12\catcode `\^12\catcode `\_12\catcode `\%12\relax}%
\providecommand \@@startlink[1]{}%
\providecommand \@@endlink[0]{}%
\providecommand \url  [0]{\begingroup\@sanitize@url \@url }%
\providecommand \@url [1]{\endgroup\@href {#1}{\urlprefix }}%
\providecommand \urlprefix  [0]{URL }%
\providecommand \Eprint [0]{\href }%
\providecommand \doibase [0]{http://dx.doi.org/}%
\providecommand \selectlanguage [0]{\@gobble}%
\providecommand \bibinfo  [0]{\@secondoftwo}%
\providecommand \bibfield  [0]{\@secondoftwo}%
\providecommand \translation [1]{[#1]}%
\providecommand \BibitemOpen [0]{}%
\providecommand \bibitemStop [0]{}%
\providecommand \bibitemNoStop [0]{.\EOS\space}%
\providecommand \EOS [0]{\spacefactor3000\relax}%
\providecommand \BibitemShut  [1]{\csname bibitem#1\endcsname}%
\let\auto@bib@innerbib\@empty
\bibitem [{\citenamefont {Liu}\ \emph {et~al.}(2024)\citenamefont {Liu}, \citenamefont {Gupta}, \citenamefont {Choi}, \citenamefont {Vituri}, \citenamefont {Stoyanov}, \citenamefont {Xiao}, \citenamefont {Wang}, \citenamefont {Zhou}, \citenamefont {Barick}, \citenamefont {Taniguchi} \emph {et~al.}}]{liu2024STM_DW}%
  \BibitemOpen
  \bibfield  {author} {\bibinfo {author} {\bibfnamefont {Yiwen}\ \bibnamefont {Liu}}, \bibinfo {author} {\bibfnamefont {Ambikesh}\ \bibnamefont {Gupta}}, \bibinfo {author} {\bibfnamefont {Youngjoon}\ \bibnamefont {Choi}}, \bibinfo {author} {\bibfnamefont {Yaar}\ \bibnamefont {Vituri}}, \bibinfo {author} {\bibfnamefont {Hari}\ \bibnamefont {Stoyanov}}, \bibinfo {author} {\bibfnamefont {Jiewen}\ \bibnamefont {Xiao}}, \bibinfo {author} {\bibfnamefont {Yanzhen}\ \bibnamefont {Wang}}, \bibinfo {author} {\bibfnamefont {Haibiao}\ \bibnamefont {Zhou}}, \bibinfo {author} {\bibfnamefont {Barun}\ \bibnamefont {Barick}}, \bibinfo {author} {\bibfnamefont {Takashi}\ \bibnamefont {Taniguchi}},  \emph {et~al.},\ }\bibfield  {title} {\enquote {\bibinfo {title} {Visualizing incommensurate inter-valley coherent states in rhombohedral trilayer graphene},}\ }\href@noop {} {\bibfield  {journal} {\bibinfo  {journal} {arXiv preprint arXiv:2411.11163}\ } (\bibinfo {year} {2024})}\BibitemShut {NoStop}%
\bibitem [{\citenamefont {Fan}\ \emph {et~al.}(2025)\citenamefont {Fan}, \citenamefont {Ou}, \citenamefont {Scholten}, \citenamefont {Li}, \citenamefont {Shang}, \citenamefont {Wang}, \citenamefont {Xu}, \citenamefont {Yang}, \citenamefont {Eremin},\ and\ \citenamefont {Wen}}]{fan2025STM_SCgap}%
  \BibitemOpen
  \bibfield  {author} {\bibinfo {author} {\bibfnamefont {Shengtai}\ \bibnamefont {Fan}}, \bibinfo {author} {\bibfnamefont {Mengjun}\ \bibnamefont {Ou}}, \bibinfo {author} {\bibfnamefont {Marius}\ \bibnamefont {Scholten}}, \bibinfo {author} {\bibfnamefont {Qing}\ \bibnamefont {Li}}, \bibinfo {author} {\bibfnamefont {Zhiyuan}\ \bibnamefont {Shang}}, \bibinfo {author} {\bibfnamefont {Yi}~\bibnamefont {Wang}}, \bibinfo {author} {\bibfnamefont {Jiasen}\ \bibnamefont {Xu}}, \bibinfo {author} {\bibfnamefont {Huan}\ \bibnamefont {Yang}}, \bibinfo {author} {\bibfnamefont {Ilya~M}\ \bibnamefont {Eremin}}, \ and\ \bibinfo {author} {\bibfnamefont {Hai-Hu}\ \bibnamefont {Wen}},\ }\bibfield  {title} {\enquote {\bibinfo {title} {Superconducting gaps revealed by stm measurements on la2prni2o7 thin films at ambient pressure},}\ }\href@noop {} {\bibfield  {journal} {\bibinfo  {journal} {arXiv preprint arXiv:2506.01788}\ } (\bibinfo {year} {2025})}\BibitemShut {NoStop}%
\bibitem [{\citenamefont {Feldman}\ \emph {et~al.}(2012)\citenamefont {Feldman}, \citenamefont {Krauss}, \citenamefont {Smet},\ and\ \citenamefont {Yacoby}}]{feldman2012SETFQHE}%
  \BibitemOpen
  \bibfield  {author} {\bibinfo {author} {\bibfnamefont {Benjamin~E}\ \bibnamefont {Feldman}}, \bibinfo {author} {\bibfnamefont {Benjamin}\ \bibnamefont {Krauss}}, \bibinfo {author} {\bibfnamefont {Jurgen~H}\ \bibnamefont {Smet}}, \ and\ \bibinfo {author} {\bibfnamefont {Amir}\ \bibnamefont {Yacoby}},\ }\bibfield  {title} {\enquote {\bibinfo {title} {Unconventional sequence of fractional quantum hall states in suspended graphene},}\ }\href@noop {} {\bibfield  {journal} {\bibinfo  {journal} {Science}\ }\textbf {\bibinfo {volume} {337}},\ \bibinfo {pages} {1196--1199} (\bibinfo {year} {2012})}\BibitemShut {NoStop}%
\bibitem [{\citenamefont {Feldman}\ \emph {et~al.}(2013)\citenamefont {Feldman}, \citenamefont {Levin}, \citenamefont {Krauss}, \citenamefont {Abanin}, \citenamefont {Halperin}, \citenamefont {Smet},\ and\ \citenamefont {Yacoby}}]{feldman2013SETFlavor}%
  \BibitemOpen
  \bibfield  {author} {\bibinfo {author} {\bibfnamefont {Benjamin~E}\ \bibnamefont {Feldman}}, \bibinfo {author} {\bibfnamefont {Andrei~J}\ \bibnamefont {Levin}}, \bibinfo {author} {\bibfnamefont {Benjamin}\ \bibnamefont {Krauss}}, \bibinfo {author} {\bibfnamefont {Dmitry~A}\ \bibnamefont {Abanin}}, \bibinfo {author} {\bibfnamefont {Bertrand~I}\ \bibnamefont {Halperin}}, \bibinfo {author} {\bibfnamefont {Jurgen~H}\ \bibnamefont {Smet}}, \ and\ \bibinfo {author} {\bibfnamefont {Amir}\ \bibnamefont {Yacoby}},\ }\bibfield  {title} {\enquote {\bibinfo {title} {Fractional quantum hall phase transitions and four-flux states in graphene},}\ }\href@noop {} {\bibfield  {journal} {\bibinfo  {journal} {Physical Review Letters}\ }\textbf {\bibinfo {volume} {111}},\ \bibinfo {pages} {076802} (\bibinfo {year} {2013})}\BibitemShut {NoStop}%
\bibitem [{\citenamefont {Grover}\ \emph {et~al.}(2022)\citenamefont {Grover}, \citenamefont {Bocarsly}, \citenamefont {Uri}, \citenamefont {Stepanov}, \citenamefont {Di~Battista}, \citenamefont {Roy}, \citenamefont {Xiao}, \citenamefont {Meltzer}, \citenamefont {Myasoedov}, \citenamefont {Pareek} \emph {et~al.}}]{grover2022SQUIDMosaic}%
  \BibitemOpen
  \bibfield  {author} {\bibinfo {author} {\bibfnamefont {Sameer}\ \bibnamefont {Grover}}, \bibinfo {author} {\bibfnamefont {Matan}\ \bibnamefont {Bocarsly}}, \bibinfo {author} {\bibfnamefont {Aviram}\ \bibnamefont {Uri}}, \bibinfo {author} {\bibfnamefont {Petr}\ \bibnamefont {Stepanov}}, \bibinfo {author} {\bibfnamefont {Giorgio}\ \bibnamefont {Di~Battista}}, \bibinfo {author} {\bibfnamefont {Indranil}\ \bibnamefont {Roy}}, \bibinfo {author} {\bibfnamefont {Jiewen}\ \bibnamefont {Xiao}}, \bibinfo {author} {\bibfnamefont {Alexander~Y}\ \bibnamefont {Meltzer}}, \bibinfo {author} {\bibfnamefont {Yuri}\ \bibnamefont {Myasoedov}}, \bibinfo {author} {\bibfnamefont {Keshav}\ \bibnamefont {Pareek}},  \emph {et~al.},\ }\bibfield  {title} {\enquote {\bibinfo {title} {Chern mosaic and berry-curvature magnetism in magic-angle graphene},}\ }\href@noop {} {\bibfield  {journal} {\bibinfo  {journal} {Nature physics}\ }\textbf {\bibinfo {volume} {18}},\ \bibinfo {pages} {885--892} (\bibinfo {year} {2022})}\BibitemShut
  {NoStop}%
\bibitem [{\citenamefont {Lai}\ \emph {et~al.}(2011)\citenamefont {Lai}, \citenamefont {Kundhikanjana}, \citenamefont {Kelly}, \citenamefont {Shen}, \citenamefont {Shabani},\ and\ \citenamefont {Shayegan}}]{Lai20112DEG}%
  \BibitemOpen
  \bibfield  {author} {\bibinfo {author} {\bibfnamefont {Keji}\ \bibnamefont {Lai}}, \bibinfo {author} {\bibfnamefont {Worasom}\ \bibnamefont {Kundhikanjana}}, \bibinfo {author} {\bibfnamefont {Michael~A}\ \bibnamefont {Kelly}}, \bibinfo {author} {\bibfnamefont {Zhi-Xun}\ \bibnamefont {Shen}}, \bibinfo {author} {\bibfnamefont {Javad}\ \bibnamefont {Shabani}}, \ and\ \bibinfo {author} {\bibfnamefont {Mansour}\ \bibnamefont {Shayegan}},\ }\bibfield  {title} {\enquote {\bibinfo {title} {Imaging of coulomb-driven quantum hall edge states},}\ }\href@noop {} {\bibfield  {journal} {\bibinfo  {journal} {Physical review letters}\ }\textbf {\bibinfo {volume} {107}},\ \bibinfo {pages} {176809} (\bibinfo {year} {2011})}\BibitemShut {NoStop}%
\bibitem [{\citenamefont {Cui}\ \emph {et~al.}(2016{\natexlab{a}})\citenamefont {Cui}, \citenamefont {Wen}, \citenamefont {Ma}, \citenamefont {Diankov}, \citenamefont {Han}, \citenamefont {Amet}, \citenamefont {Taniguchi}, \citenamefont {Watanabe}, \citenamefont {Goldhaber-Gordon}, \citenamefont {Dean} \emph {et~al.}}]{Cui2016Gr}%
  \BibitemOpen
  \bibfield  {author} {\bibinfo {author} {\bibfnamefont {Yong-Tao}\ \bibnamefont {Cui}}, \bibinfo {author} {\bibfnamefont {Bo}~\bibnamefont {Wen}}, \bibinfo {author} {\bibfnamefont {Eric~Y}\ \bibnamefont {Ma}}, \bibinfo {author} {\bibfnamefont {Georgi}\ \bibnamefont {Diankov}}, \bibinfo {author} {\bibfnamefont {Zheng}\ \bibnamefont {Han}}, \bibinfo {author} {\bibfnamefont {Francois}\ \bibnamefont {Amet}}, \bibinfo {author} {\bibfnamefont {Takashi}\ \bibnamefont {Taniguchi}}, \bibinfo {author} {\bibfnamefont {Kenji}\ \bibnamefont {Watanabe}}, \bibinfo {author} {\bibfnamefont {David}\ \bibnamefont {Goldhaber-Gordon}}, \bibinfo {author} {\bibfnamefont {Cory~R}\ \bibnamefont {Dean}},  \emph {et~al.},\ }\bibfield  {title} {\enquote {\bibinfo {title} {Unconventional correlation between quantum hall transport quantization and bulk state filling in gated graphene devices},}\ }\href@noop {} {\bibfield  {journal} {\bibinfo  {journal} {Physical Review Letters}\ }\textbf {\bibinfo {volume} {117}},\ \bibinfo {pages}
  {186601} (\bibinfo {year} {2016}{\natexlab{a}})}\BibitemShut {NoStop}%
\bibitem [{\citenamefont {Shi}\ \emph {et~al.}(2019)\citenamefont {Shi}, \citenamefont {Kahn}, \citenamefont {Niu}, \citenamefont {Fei}, \citenamefont {Sun}, \citenamefont {Cai}, \citenamefont {Francisco}, \citenamefont {Wu}, \citenamefont {Shen}, \citenamefont {Xu} \emph {et~al.}}]{Shi2019WTe2}%
  \BibitemOpen
  \bibfield  {author} {\bibinfo {author} {\bibfnamefont {Yanmeng}\ \bibnamefont {Shi}}, \bibinfo {author} {\bibfnamefont {Joshua}\ \bibnamefont {Kahn}}, \bibinfo {author} {\bibfnamefont {Ben}\ \bibnamefont {Niu}}, \bibinfo {author} {\bibfnamefont {Zaiyao}\ \bibnamefont {Fei}}, \bibinfo {author} {\bibfnamefont {Bosong}\ \bibnamefont {Sun}}, \bibinfo {author} {\bibfnamefont {Xinghan}\ \bibnamefont {Cai}}, \bibinfo {author} {\bibfnamefont {Brian~A}\ \bibnamefont {Francisco}}, \bibinfo {author} {\bibfnamefont {Di}~\bibnamefont {Wu}}, \bibinfo {author} {\bibfnamefont {Zhi-Xun}\ \bibnamefont {Shen}}, \bibinfo {author} {\bibfnamefont {Xiaodong}\ \bibnamefont {Xu}},  \emph {et~al.},\ }\bibfield  {title} {\enquote {\bibinfo {title} {Imaging quantum spin hall edges in monolayer wte2},}\ }\href@noop {} {\bibfield  {journal} {\bibinfo  {journal} {Science advances}\ }\textbf {\bibinfo {volume} {5}},\ \bibinfo {pages} {eaat8799} (\bibinfo {year} {2019})}\BibitemShut {NoStop}%
\bibitem [{\citenamefont {Allen}\ \emph {et~al.}(2019)\citenamefont {Allen}, \citenamefont {Cui}, \citenamefont {Yue~Ma}, \citenamefont {Mogi}, \citenamefont {Kawamura}, \citenamefont {Fulga}, \citenamefont {Goldhaber-Gordon}, \citenamefont {Tokura},\ and\ \citenamefont {Shen}}]{allen2019MIM}%
  \BibitemOpen
  \bibfield  {author} {\bibinfo {author} {\bibfnamefont {Monica}\ \bibnamefont {Allen}}, \bibinfo {author} {\bibfnamefont {Yongtao}\ \bibnamefont {Cui}}, \bibinfo {author} {\bibfnamefont {Eric}\ \bibnamefont {Yue~Ma}}, \bibinfo {author} {\bibfnamefont {Masataka}\ \bibnamefont {Mogi}}, \bibinfo {author} {\bibfnamefont {Minoru}\ \bibnamefont {Kawamura}}, \bibinfo {author} {\bibfnamefont {Ion~Cosma}\ \bibnamefont {Fulga}}, \bibinfo {author} {\bibfnamefont {David}\ \bibnamefont {Goldhaber-Gordon}}, \bibinfo {author} {\bibfnamefont {Yoshinori}\ \bibnamefont {Tokura}}, \ and\ \bibinfo {author} {\bibfnamefont {Zhi-Xun}\ \bibnamefont {Shen}},\ }\bibfield  {title} {\enquote {\bibinfo {title} {Visualization of an axion insulating state at the transition between 2 chiral quantum anomalous hall states},}\ }\href@noop {} {\bibfield  {journal} {\bibinfo  {journal} {Proceedings of the National Academy of Sciences}\ }\textbf {\bibinfo {volume} {116}},\ \bibinfo {pages} {14511--14515} (\bibinfo {year} {2019})}\BibitemShut
  {NoStop}%
\bibitem [{\citenamefont {Martin}\ \emph {et~al.}(1988)\citenamefont {Martin}, \citenamefont {Abraham},\ and\ \citenamefont {Wickramasinghe}}]{martin1988EFM_surface}%
  \BibitemOpen
  \bibfield  {author} {\bibinfo {author} {\bibfnamefont {Yves}\ \bibnamefont {Martin}}, \bibinfo {author} {\bibfnamefont {David~W}\ \bibnamefont {Abraham}}, \ and\ \bibinfo {author} {\bibfnamefont {H~Kumar}\ \bibnamefont {Wickramasinghe}},\ }\bibfield  {title} {\enquote {\bibinfo {title} {High-resolution capacitance measurement and potentiometry by force microscopy},}\ }\href@noop {} {\bibfield  {journal} {\bibinfo  {journal} {Applied Physics Letters}\ }\textbf {\bibinfo {volume} {52}},\ \bibinfo {pages} {1103--1105} (\bibinfo {year} {1988})}\BibitemShut {NoStop}%
\bibitem [{\citenamefont {Nonnenmacher}\ \emph {et~al.}(1991)\citenamefont {Nonnenmacher}, \citenamefont {o’Boyle},\ and\ \citenamefont {Wickramasinghe}}]{nonnenmacher1991KPFM_surface}%
  \BibitemOpen
  \bibfield  {author} {\bibinfo {author} {\bibfnamefont {Manuel}\ \bibnamefont {Nonnenmacher}}, \bibinfo {author} {\bibfnamefont {MP}~\bibnamefont {o’Boyle}}, \ and\ \bibinfo {author} {\bibfnamefont {H~Kumar}\ \bibnamefont {Wickramasinghe}},\ }\bibfield  {title} {\enquote {\bibinfo {title} {Kelvin probe force microscopy},}\ }\href@noop {} {\bibfield  {journal} {\bibinfo  {journal} {Applied physics letters}\ }\textbf {\bibinfo {volume} {58}},\ \bibinfo {pages} {2921--2923} (\bibinfo {year} {1991})}\BibitemShut {NoStop}%
\bibitem [{\citenamefont {Yoo}\ \emph {et~al.}(1997)\citenamefont {Yoo}, \citenamefont {Fulton}, \citenamefont {Hess}, \citenamefont {Willett}, \citenamefont {Dunkleberger}, \citenamefont {Chichester}, \citenamefont {Pfeiffer},\ and\ \citenamefont {West}}]{yoo1997SSET_surface}%
  \BibitemOpen
  \bibfield  {author} {\bibinfo {author} {\bibfnamefont {MJ}~\bibnamefont {Yoo}}, \bibinfo {author} {\bibfnamefont {TA}~\bibnamefont {Fulton}}, \bibinfo {author} {\bibfnamefont {HF}~\bibnamefont {Hess}}, \bibinfo {author} {\bibfnamefont {RL}~\bibnamefont {Willett}}, \bibinfo {author} {\bibfnamefont {LN}~\bibnamefont {Dunkleberger}}, \bibinfo {author} {\bibfnamefont {RJ}~\bibnamefont {Chichester}}, \bibinfo {author} {\bibfnamefont {LN}~\bibnamefont {Pfeiffer}}, \ and\ \bibinfo {author} {\bibfnamefont {KW}~\bibnamefont {West}},\ }\bibfield  {title} {\enquote {\bibinfo {title} {Scanning single-electron transistor microscopy: Imaging individual charges},}\ }\href@noop {} {\bibfield  {journal} {\bibinfo  {journal} {Science}\ }\textbf {\bibinfo {volume} {276}},\ \bibinfo {pages} {579--582} (\bibinfo {year} {1997})}\BibitemShut {NoStop}%
\bibitem [{\citenamefont {Zhang}\ \emph {et~al.}(2009)\citenamefont {Zhang}, \citenamefont {Tang}, \citenamefont {Girit}, \citenamefont {Hao}, \citenamefont {Martin}, \citenamefont {Zettl}, \citenamefont {Crommie}, \citenamefont {Shen},\ and\ \citenamefont {Wang}}]{Zhang2009BLG}%
  \BibitemOpen
  \bibfield  {author} {\bibinfo {author} {\bibfnamefont {Yuanbo}\ \bibnamefont {Zhang}}, \bibinfo {author} {\bibfnamefont {Tsung-Ta}\ \bibnamefont {Tang}}, \bibinfo {author} {\bibfnamefont {Caglar}\ \bibnamefont {Girit}}, \bibinfo {author} {\bibfnamefont {Zhao}\ \bibnamefont {Hao}}, \bibinfo {author} {\bibfnamefont {Michael~C}\ \bibnamefont {Martin}}, \bibinfo {author} {\bibfnamefont {Alex}\ \bibnamefont {Zettl}}, \bibinfo {author} {\bibfnamefont {Michael~F}\ \bibnamefont {Crommie}}, \bibinfo {author} {\bibfnamefont {Y~Ron}\ \bibnamefont {Shen}}, \ and\ \bibinfo {author} {\bibfnamefont {Feng}\ \bibnamefont {Wang}},\ }\bibfield  {title} {\enquote {\bibinfo {title} {Direct observation of a widely tunable bandgap in bilayer graphene},}\ }\href@noop {} {\bibfield  {journal} {\bibinfo  {journal} {Nature}\ }\textbf {\bibinfo {volume} {459}},\ \bibinfo {pages} {820--823} (\bibinfo {year} {2009})}\BibitemShut {NoStop}%
\bibitem [{\citenamefont {Cao}\ \emph {et~al.}(2018)\citenamefont {Cao}, \citenamefont {Fatemi}, \citenamefont {Demir}, \citenamefont {Fang}, \citenamefont {Tomarken}, \citenamefont {Luo}, \citenamefont {Sanchez-Yamagishi}, \citenamefont {Watanabe}, \citenamefont {Taniguchi}, \citenamefont {Kaxiras} \emph {et~al.}}]{Cao2018MATBG}%
  \BibitemOpen
  \bibfield  {author} {\bibinfo {author} {\bibfnamefont {Yuan}\ \bibnamefont {Cao}}, \bibinfo {author} {\bibfnamefont {Valla}\ \bibnamefont {Fatemi}}, \bibinfo {author} {\bibfnamefont {Ahmet}\ \bibnamefont {Demir}}, \bibinfo {author} {\bibfnamefont {Shiang}\ \bibnamefont {Fang}}, \bibinfo {author} {\bibfnamefont {Spencer~L}\ \bibnamefont {Tomarken}}, \bibinfo {author} {\bibfnamefont {Jason~Y}\ \bibnamefont {Luo}}, \bibinfo {author} {\bibfnamefont {Javier~D}\ \bibnamefont {Sanchez-Yamagishi}}, \bibinfo {author} {\bibfnamefont {Kenji}\ \bibnamefont {Watanabe}}, \bibinfo {author} {\bibfnamefont {Takashi}\ \bibnamefont {Taniguchi}}, \bibinfo {author} {\bibfnamefont {Efthimios}\ \bibnamefont {Kaxiras}},  \emph {et~al.},\ }\bibfield  {title} {\enquote {\bibinfo {title} {Correlated insulator behaviour at half-filling in magic-angle graphene superlattices},}\ }\href@noop {} {\bibfield  {journal} {\bibinfo  {journal} {Nature}\ }\textbf {\bibinfo {volume} {556}},\ \bibinfo {pages} {80--84} (\bibinfo {year}
  {2018})}\BibitemShut {NoStop}%
\bibitem [{\citenamefont {Wang}\ \emph {et~al.}(2020)\citenamefont {Wang}, \citenamefont {Shih}, \citenamefont {Ghiotto}, \citenamefont {Xian}, \citenamefont {Rhodes}, \citenamefont {Tan}, \citenamefont {Claassen}, \citenamefont {Kennes}, \citenamefont {Bai}, \citenamefont {Kim} \emph {et~al.}}]{Wang2020WSe2WSe2}%
  \BibitemOpen
  \bibfield  {author} {\bibinfo {author} {\bibfnamefont {Lei}\ \bibnamefont {Wang}}, \bibinfo {author} {\bibfnamefont {En-Min}\ \bibnamefont {Shih}}, \bibinfo {author} {\bibfnamefont {Augusto}\ \bibnamefont {Ghiotto}}, \bibinfo {author} {\bibfnamefont {Lede}\ \bibnamefont {Xian}}, \bibinfo {author} {\bibfnamefont {Daniel~A}\ \bibnamefont {Rhodes}}, \bibinfo {author} {\bibfnamefont {Cheng}\ \bibnamefont {Tan}}, \bibinfo {author} {\bibfnamefont {Martin}\ \bibnamefont {Claassen}}, \bibinfo {author} {\bibfnamefont {Dante~M}\ \bibnamefont {Kennes}}, \bibinfo {author} {\bibfnamefont {Yusong}\ \bibnamefont {Bai}}, \bibinfo {author} {\bibfnamefont {Bumho}\ \bibnamefont {Kim}},  \emph {et~al.},\ }\bibfield  {title} {\enquote {\bibinfo {title} {Correlated electronic phases in twisted bilayer transition metal dichalcogenides},}\ }\href@noop {} {\bibfield  {journal} {\bibinfo  {journal} {Nature materials}\ }\textbf {\bibinfo {volume} {19}},\ \bibinfo {pages} {861--866} (\bibinfo {year} {2020})}\BibitemShut {NoStop}%
\bibitem [{\citenamefont {Li}\ \emph {et~al.}(2021{\natexlab{a}})\citenamefont {Li}, \citenamefont {Zhu}, \citenamefont {Tang}, \citenamefont {Watanabe}, \citenamefont {Taniguchi}, \citenamefont {Elser}, \citenamefont {Shan},\ and\ \citenamefont {Mak}}]{Li2021MoSe2WS2}%
  \BibitemOpen
  \bibfield  {author} {\bibinfo {author} {\bibfnamefont {Tingxin}\ \bibnamefont {Li}}, \bibinfo {author} {\bibfnamefont {Jiacheng}\ \bibnamefont {Zhu}}, \bibinfo {author} {\bibfnamefont {Yanhao}\ \bibnamefont {Tang}}, \bibinfo {author} {\bibfnamefont {Kenji}\ \bibnamefont {Watanabe}}, \bibinfo {author} {\bibfnamefont {Takashi}\ \bibnamefont {Taniguchi}}, \bibinfo {author} {\bibfnamefont {Veit}\ \bibnamefont {Elser}}, \bibinfo {author} {\bibfnamefont {Jie}\ \bibnamefont {Shan}}, \ and\ \bibinfo {author} {\bibfnamefont {Kin~Fai}\ \bibnamefont {Mak}},\ }\bibfield  {title} {\enquote {\bibinfo {title} {Charge-order-enhanced capacitance in semiconductor moir{\'e} superlattices},}\ }\href@noop {} {\bibfield  {journal} {\bibinfo  {journal} {Nature Nanotechnology}\ }\textbf {\bibinfo {volume} {16}},\ \bibinfo {pages} {1068--1072} (\bibinfo {year} {2021}{\natexlab{a}})}\BibitemShut {NoStop}%
\bibitem [{\citenamefont {Cai}\ \emph {et~al.}(2023)\citenamefont {Cai}, \citenamefont {Anderson}, \citenamefont {Wang}, \citenamefont {Zhang}, \citenamefont {Liu}, \citenamefont {Holtzmann}, \citenamefont {Zhang}, \citenamefont {Fan}, \citenamefont {Taniguchi}, \citenamefont {Watanabe} \emph {et~al.}}]{Cai2023MoTe2}%
  \BibitemOpen
  \bibfield  {author} {\bibinfo {author} {\bibfnamefont {Jiaqi}\ \bibnamefont {Cai}}, \bibinfo {author} {\bibfnamefont {Eric}\ \bibnamefont {Anderson}}, \bibinfo {author} {\bibfnamefont {Chong}\ \bibnamefont {Wang}}, \bibinfo {author} {\bibfnamefont {Xiaowei}\ \bibnamefont {Zhang}}, \bibinfo {author} {\bibfnamefont {Xiaoyu}\ \bibnamefont {Liu}}, \bibinfo {author} {\bibfnamefont {William}\ \bibnamefont {Holtzmann}}, \bibinfo {author} {\bibfnamefont {Yinong}\ \bibnamefont {Zhang}}, \bibinfo {author} {\bibfnamefont {Fengren}\ \bibnamefont {Fan}}, \bibinfo {author} {\bibfnamefont {Takashi}\ \bibnamefont {Taniguchi}}, \bibinfo {author} {\bibfnamefont {Kenji}\ \bibnamefont {Watanabe}},  \emph {et~al.},\ }\bibfield  {title} {\enquote {\bibinfo {title} {Signatures of fractional quantum anomalous hall states in twisted mote2},}\ }\href@noop {} {\bibfield  {journal} {\bibinfo  {journal} {Nature}\ }\textbf {\bibinfo {volume} {622}},\ \bibinfo {pages} {63--68} (\bibinfo {year} {2023})}\BibitemShut {NoStop}%
\bibitem [{\citenamefont {Park}\ \emph {et~al.}(2024)\citenamefont {Park}, \citenamefont {Cai}, \citenamefont {Anderson}, \citenamefont {Zhang}, \citenamefont {Liu}, \citenamefont {Holtzmann}, \citenamefont {Li}, \citenamefont {Wang}, \citenamefont {Hu}, \citenamefont {Zhao} \emph {et~al.}}]{Park2024MoTe2}%
  \BibitemOpen
  \bibfield  {author} {\bibinfo {author} {\bibfnamefont {Heonjoon}\ \bibnamefont {Park}}, \bibinfo {author} {\bibfnamefont {Jiaqi}\ \bibnamefont {Cai}}, \bibinfo {author} {\bibfnamefont {Eric}\ \bibnamefont {Anderson}}, \bibinfo {author} {\bibfnamefont {Xiao-Wei}\ \bibnamefont {Zhang}}, \bibinfo {author} {\bibfnamefont {Xiaoyu}\ \bibnamefont {Liu}}, \bibinfo {author} {\bibfnamefont {William}\ \bibnamefont {Holtzmann}}, \bibinfo {author} {\bibfnamefont {Weijie}\ \bibnamefont {Li}}, \bibinfo {author} {\bibfnamefont {Chong}\ \bibnamefont {Wang}}, \bibinfo {author} {\bibfnamefont {Chaowei}\ \bibnamefont {Hu}}, \bibinfo {author} {\bibfnamefont {Yuzhou}\ \bibnamefont {Zhao}},  \emph {et~al.},\ }\bibfield  {title} {\enquote {\bibinfo {title} {Ferromagnetism and topology of the higher flat band in a fractional chern insulator},}\ }\href@noop {} {\bibfield  {journal} {\bibinfo  {journal} {arXiv preprint arXiv:2406.09591}\ } (\bibinfo {year} {2024})}\BibitemShut {NoStop}%
\bibitem [{\citenamefont {Ji}\ \emph {et~al.}(2024)\citenamefont {Ji}, \citenamefont {Park}, \citenamefont {Barber}, \citenamefont {Hu}, \citenamefont {Watanabe}, \citenamefont {Taniguchi}, \citenamefont {Chu}, \citenamefont {Xu},\ and\ \citenamefont {Shen}}]{Ji2024MoTe2}%
  \BibitemOpen
  \bibfield  {author} {\bibinfo {author} {\bibfnamefont {Zhurun}\ \bibnamefont {Ji}}, \bibinfo {author} {\bibfnamefont {Heonjoon}\ \bibnamefont {Park}}, \bibinfo {author} {\bibfnamefont {Mark~E}\ \bibnamefont {Barber}}, \bibinfo {author} {\bibfnamefont {Chaowei}\ \bibnamefont {Hu}}, \bibinfo {author} {\bibfnamefont {Kenji}\ \bibnamefont {Watanabe}}, \bibinfo {author} {\bibfnamefont {Takashi}\ \bibnamefont {Taniguchi}}, \bibinfo {author} {\bibfnamefont {Jiun-Haw}\ \bibnamefont {Chu}}, \bibinfo {author} {\bibfnamefont {Xiaodong}\ \bibnamefont {Xu}}, \ and\ \bibinfo {author} {\bibfnamefont {Zhi-Xun}\ \bibnamefont {Shen}},\ }\bibfield  {title} {\enquote {\bibinfo {title} {Local probe of bulk and edge states in a fractional chern insulator},}\ }\href@noop {} {\bibfield  {journal} {\bibinfo  {journal} {Nature}\ }\textbf {\bibinfo {volume} {635}},\ \bibinfo {pages} {578--583} (\bibinfo {year} {2024})}\BibitemShut {NoStop}%
\bibitem [{\citenamefont {Xu}\ \emph {et~al.}(2020)\citenamefont {Xu}, \citenamefont {Liu}, \citenamefont {Rhodes}, \citenamefont {Watanabe}, \citenamefont {Taniguchi}, \citenamefont {Hone}, \citenamefont {Elser}, \citenamefont {Mak},\ and\ \citenamefont {Shan}}]{Xu2020WSe2WS2}%
  \BibitemOpen
  \bibfield  {author} {\bibinfo {author} {\bibfnamefont {Yang}\ \bibnamefont {Xu}}, \bibinfo {author} {\bibfnamefont {Song}\ \bibnamefont {Liu}}, \bibinfo {author} {\bibfnamefont {Daniel~A}\ \bibnamefont {Rhodes}}, \bibinfo {author} {\bibfnamefont {Kenji}\ \bibnamefont {Watanabe}}, \bibinfo {author} {\bibfnamefont {Takashi}\ \bibnamefont {Taniguchi}}, \bibinfo {author} {\bibfnamefont {James}\ \bibnamefont {Hone}}, \bibinfo {author} {\bibfnamefont {Veit}\ \bibnamefont {Elser}}, \bibinfo {author} {\bibfnamefont {Kin~Fai}\ \bibnamefont {Mak}}, \ and\ \bibinfo {author} {\bibfnamefont {Jie}\ \bibnamefont {Shan}},\ }\bibfield  {title} {\enquote {\bibinfo {title} {Correlated insulating states at fractional fillings of moir{\'e} superlattices},}\ }\href@noop {} {\bibfield  {journal} {\bibinfo  {journal} {Nature}\ }\textbf {\bibinfo {volume} {587}},\ \bibinfo {pages} {214--218} (\bibinfo {year} {2020})}\BibitemShut {NoStop}%
\bibitem [{\citenamefont {Zeng}\ \emph {et~al.}(2023)\citenamefont {Zeng}, \citenamefont {Xia}, \citenamefont {Kang}, \citenamefont {Zhu}, \citenamefont {Kn{\"u}ppel}, \citenamefont {Vaswani}, \citenamefont {Watanabe}, \citenamefont {Taniguchi}, \citenamefont {Mak},\ and\ \citenamefont {Shan}}]{Zeng2023MoTe2}%
  \BibitemOpen
  \bibfield  {author} {\bibinfo {author} {\bibfnamefont {Yihang}\ \bibnamefont {Zeng}}, \bibinfo {author} {\bibfnamefont {Zhengchao}\ \bibnamefont {Xia}}, \bibinfo {author} {\bibfnamefont {Kaifei}\ \bibnamefont {Kang}}, \bibinfo {author} {\bibfnamefont {Jiacheng}\ \bibnamefont {Zhu}}, \bibinfo {author} {\bibfnamefont {Patrick}\ \bibnamefont {Kn{\"u}ppel}}, \bibinfo {author} {\bibfnamefont {Chirag}\ \bibnamefont {Vaswani}}, \bibinfo {author} {\bibfnamefont {Kenji}\ \bibnamefont {Watanabe}}, \bibinfo {author} {\bibfnamefont {Takashi}\ \bibnamefont {Taniguchi}}, \bibinfo {author} {\bibfnamefont {Kin~Fai}\ \bibnamefont {Mak}}, \ and\ \bibinfo {author} {\bibfnamefont {Jie}\ \bibnamefont {Shan}},\ }\bibfield  {title} {\enquote {\bibinfo {title} {Thermodynamic evidence of fractional chern insulator in moir{\'e} mote2},}\ }\href@noop {} {\bibfield  {journal} {\bibinfo  {journal} {Nature}\ }\textbf {\bibinfo {volume} {622}},\ \bibinfo {pages} {69--73} (\bibinfo {year} {2023})}\BibitemShut {NoStop}%
\bibitem [{\citenamefont {Xia}\ \emph {et~al.}(2024)\citenamefont {Xia}, \citenamefont {Zeng}, \citenamefont {Shen}, \citenamefont {Dery}, \citenamefont {Watanabe}, \citenamefont {Taniguchi}, \citenamefont {Shan},\ and\ \citenamefont {Mak}}]{Xia2024Sensor}%
  \BibitemOpen
  \bibfield  {author} {\bibinfo {author} {\bibfnamefont {Zhengchao}\ \bibnamefont {Xia}}, \bibinfo {author} {\bibfnamefont {Yihang}\ \bibnamefont {Zeng}}, \bibinfo {author} {\bibfnamefont {Bowen}\ \bibnamefont {Shen}}, \bibinfo {author} {\bibfnamefont {Roei}\ \bibnamefont {Dery}}, \bibinfo {author} {\bibfnamefont {Kenji}\ \bibnamefont {Watanabe}}, \bibinfo {author} {\bibfnamefont {Takashi}\ \bibnamefont {Taniguchi}}, \bibinfo {author} {\bibfnamefont {Jie}\ \bibnamefont {Shan}}, \ and\ \bibinfo {author} {\bibfnamefont {Kin~Fai}\ \bibnamefont {Mak}},\ }\bibfield  {title} {\enquote {\bibinfo {title} {Optical readout of the chemical potential of two-dimensional electrons},}\ }\href@noop {} {\bibfield  {journal} {\bibinfo  {journal} {Nature Photonics}\ }\textbf {\bibinfo {volume} {18}},\ \bibinfo {pages} {344--349} (\bibinfo {year} {2024})}\BibitemShut {NoStop}%
\bibitem [{\citenamefont {Eich}\ \emph {et~al.}(2018)\citenamefont {Eich}, \citenamefont {Herman}, \citenamefont {Pisoni}, \citenamefont {Overweg}, \citenamefont {Kurzmann}, \citenamefont {Lee}, \citenamefont {Rickhaus}, \citenamefont {Watanabe}, \citenamefont {Taniguchi}, \citenamefont {Sigrist}, \citenamefont {Ihn},\ and\ \citenamefont {Ensslin}}]{Enssline2018BGQD}%
  \BibitemOpen
  \bibfield  {author} {\bibinfo {author} {\bibfnamefont {Marius}\ \bibnamefont {Eich}}, \bibinfo {author} {\bibfnamefont {Franti\ifmmode \check{s}\else~\v{s}\fi{}ek}\ \bibnamefont {Herman}}, \bibinfo {author} {\bibfnamefont {Riccardo}\ \bibnamefont {Pisoni}}, \bibinfo {author} {\bibfnamefont {Hiske}\ \bibnamefont {Overweg}}, \bibinfo {author} {\bibfnamefont {Annika}\ \bibnamefont {Kurzmann}}, \bibinfo {author} {\bibfnamefont {Yongjin}\ \bibnamefont {Lee}}, \bibinfo {author} {\bibfnamefont {Peter}\ \bibnamefont {Rickhaus}}, \bibinfo {author} {\bibfnamefont {Kenji}\ \bibnamefont {Watanabe}}, \bibinfo {author} {\bibfnamefont {Takashi}\ \bibnamefont {Taniguchi}}, \bibinfo {author} {\bibfnamefont {Manfred}\ \bibnamefont {Sigrist}}, \bibinfo {author} {\bibfnamefont {Thomas}\ \bibnamefont {Ihn}}, \ and\ \bibinfo {author} {\bibfnamefont {Klaus}\ \bibnamefont {Ensslin}},\ }\bibfield  {title} {\enquote {\bibinfo {title} {Spin and valley states in gate-defined bilayer graphene quantum dots},}\ }\href@noop {} {\bibfield
  {journal} {\bibinfo  {journal} {Phys. Rev. X}\ }\textbf {\bibinfo {volume} {8}},\ \bibinfo {pages} {031023} (\bibinfo {year} {2018})}\BibitemShut {NoStop}%
\bibitem [{\citenamefont {Goossens}\ \emph {et~al.}(2012{\natexlab{a}})\citenamefont {Goossens}, \citenamefont {Driessen}, \citenamefont {Baart}, \citenamefont {Watanabe}, \citenamefont {Taniguchi},\ and\ \citenamefont {Vandersypen}}]{Vandersypen2012BGhBN}%
  \BibitemOpen
  \bibfield  {author} {\bibinfo {author} {\bibfnamefont {Augustinus (Stijn)~M.}\ \bibnamefont {Goossens}}, \bibinfo {author} {\bibfnamefont {Stefanie C.~M.}\ \bibnamefont {Driessen}}, \bibinfo {author} {\bibfnamefont {Tim~A.}\ \bibnamefont {Baart}}, \bibinfo {author} {\bibfnamefont {Kenji}\ \bibnamefont {Watanabe}}, \bibinfo {author} {\bibfnamefont {Takashi}\ \bibnamefont {Taniguchi}}, \ and\ \bibinfo {author} {\bibfnamefont {Lieven M.~K.}\ \bibnamefont {Vandersypen}},\ }\bibfield  {title} {\enquote {\bibinfo {title} {Gate-defined confinement in bilayer graphene-hexagonal boron nitride hybrid devices},}\ }\href@noop {} {\bibfield  {journal} {\bibinfo  {journal} {Nano Letters}\ }\textbf {\bibinfo {volume} {12}},\ \bibinfo {pages} {4656--4660} (\bibinfo {year} {2012}{\natexlab{a}})}\BibitemShut {NoStop}%
\bibitem [{\citenamefont {Allen}\ \emph {et~al.}(2012)\citenamefont {Allen}, \citenamefont {Martin},\ and\ \citenamefont {Yacoby}}]{Yacoby2012SBG}%
  \BibitemOpen
  \bibfield  {author} {\bibinfo {author} {\bibfnamefont {Monica}\ \bibnamefont {Allen}}, \bibinfo {author} {\bibfnamefont {Jens}\ \bibnamefont {Martin}}, \ and\ \bibinfo {author} {\bibfnamefont {Amir}\ \bibnamefont {Yacoby}},\ }\bibfield  {title} {\enquote {\bibinfo {title} {Gate defined quantum confinement in suspended bilayer graphene},}\ }\href@noop {} {\bibfield  {journal} {\bibinfo  {journal} {Nature communications}\ }\textbf {\bibinfo {volume} {3}},\ \bibinfo {pages} {934} (\bibinfo {year} {2012})}\BibitemShut {NoStop}%
\bibitem [{\citenamefont {Kim}\ \emph {et~al.}(2023)\citenamefont {Kim}, \citenamefont {Choi}, \citenamefont {Lantagne-Hurtubise}, \citenamefont {Lewandowski}, \citenamefont {Thomson}, \citenamefont {Kong}, \citenamefont {Zhou}, \citenamefont {Baum}, \citenamefont {Zhang}, \citenamefont {Holleis}, \citenamefont {Watanabe}, \citenamefont {Taniguchi}, \citenamefont {Young}, \citenamefont {Alicea},\ and\ \citenamefont {Nadj-Perge}}]{Nadj-Perge2023TLG}%
  \BibitemOpen
  \bibfield  {author} {\bibinfo {author} {\bibfnamefont {Hyunjin}\ \bibnamefont {Kim}}, \bibinfo {author} {\bibfnamefont {Youngjoon}\ \bibnamefont {Choi}}, \bibinfo {author} {\bibfnamefont {Étienne}\ \bibnamefont {Lantagne-Hurtubise}}, \bibinfo {author} {\bibfnamefont {Cyprian}\ \bibnamefont {Lewandowski}}, \bibinfo {author} {\bibfnamefont {Alex}\ \bibnamefont {Thomson}}, \bibinfo {author} {\bibfnamefont {Lingyuan}\ \bibnamefont {Kong}}, \bibinfo {author} {\bibfnamefont {Haoxin}\ \bibnamefont {Zhou}}, \bibinfo {author} {\bibfnamefont {Eli}\ \bibnamefont {Baum}}, \bibinfo {author} {\bibfnamefont {Yiran}\ \bibnamefont {Zhang}}, \bibinfo {author} {\bibfnamefont {Ludwig}\ \bibnamefont {Holleis}}, \bibinfo {author} {\bibfnamefont {Kenji}\ \bibnamefont {Watanabe}}, \bibinfo {author} {\bibfnamefont {Takashi}\ \bibnamefont {Taniguchi}}, \bibinfo {author} {\bibfnamefont {Andrea~F.}\ \bibnamefont {Young}}, \bibinfo {author} {\bibfnamefont {Jason}\ \bibnamefont {Alicea}}, \ and\ \bibinfo {author} {\bibfnamefont
  {Stevan}\ \bibnamefont {Nadj-Perge}},\ }\bibfield  {title} {\enquote {\bibinfo {title} {Imaging inter-valley coherent order in magic-angle twisted trilayer graphene},}\ }\href@noop {} {\bibfield  {journal} {\bibinfo  {journal} {Nature}\ }\textbf {\bibinfo {volume} {623}},\ \bibinfo {pages} {942–948} (\bibinfo {year} {2023})}\BibitemShut {NoStop}%
\bibitem [{\citenamefont {Han}\ \emph {et~al.}(2024)\citenamefont {Han}, \citenamefont {Lu}, \citenamefont {Scuri}, \citenamefont {Sung}, \citenamefont {Wang}, \citenamefont {Han}, \citenamefont {Watanabe}, \citenamefont {Taniguchi}, \citenamefont {Park},\ and\ \citenamefont {Ju}}]{Ju2024CIPG}%
  \BibitemOpen
  \bibfield  {author} {\bibinfo {author} {\bibfnamefont {Tonghang}\ \bibnamefont {Han}}, \bibinfo {author} {\bibfnamefont {Zhengguang}\ \bibnamefont {Lu}}, \bibinfo {author} {\bibfnamefont {Giovanni}\ \bibnamefont {Scuri}}, \bibinfo {author} {\bibfnamefont {Jiho}\ \bibnamefont {Sung}}, \bibinfo {author} {\bibfnamefont {Jue}\ \bibnamefont {Wang}}, \bibinfo {author} {\bibfnamefont {Tianyi}\ \bibnamefont {Han}}, \bibinfo {author} {\bibfnamefont {Kenji}\ \bibnamefont {Watanabe}}, \bibinfo {author} {\bibfnamefont {Takashi}\ \bibnamefont {Taniguchi}}, \bibinfo {author} {\bibfnamefont {Hongkun}\ \bibnamefont {Park}}, \ and\ \bibinfo {author} {\bibfnamefont {Long}\ \bibnamefont {Ju}},\ }\bibfield  {title} {\enquote {\bibinfo {title} {Correlated insulator and chern insulators in pentalayer rhombohedral-stacked graphene},}\ }\href@noop {} {\bibfield  {journal} {\bibinfo  {journal} {Nature Nanotechnology}\ }\textbf {\bibinfo {volume} {19}},\ \bibinfo {pages} {181–187} (\bibinfo {year} {2024})}\BibitemShut {NoStop}%
\bibitem [{\citenamefont {Lu}\ \emph {et~al.}(2024)\citenamefont {Lu}, \citenamefont {Han}, \citenamefont {Yao}, \citenamefont {Reddy}, \citenamefont {Yang}, \citenamefont {Seo}, \citenamefont {Watanabe}, \citenamefont {Taniguchi},\ and\ \citenamefont {Ju}}]{Ju2024FQAHMG}%
  \BibitemOpen
  \bibfield  {author} {\bibinfo {author} {\bibfnamefont {Zhengguang}\ \bibnamefont {Lu}}, \bibinfo {author} {\bibfnamefont {Tonghang}\ \bibnamefont {Han}}, \bibinfo {author} {\bibfnamefont {Yuxuan}\ \bibnamefont {Yao}}, \bibinfo {author} {\bibfnamefont {Aidan}\ \bibnamefont {Reddy}}, \bibinfo {author} {\bibfnamefont {Jixiang}\ \bibnamefont {Yang}}, \bibinfo {author} {\bibfnamefont {Junseok}\ \bibnamefont {Seo}}, \bibinfo {author} {\bibfnamefont {Kenji}\ \bibnamefont {Watanabe}}, \bibinfo {author} {\bibfnamefont {Takashi}\ \bibnamefont {Taniguchi}}, \ and\ \bibinfo {author} {\bibfnamefont {Long}\ \bibnamefont {Ju}},\ }\bibfield  {title} {\enquote {\bibinfo {title} {Fractional quantum anomalous hall effect in multilayer graphene},}\ }\href@noop {} {\bibfield  {journal} {\bibinfo  {journal} {Nature}\ }\textbf {\bibinfo {volume} {626}},\ \bibinfo {pages} {759--764} (\bibinfo {year} {2024})}\BibitemShut {NoStop}%
\bibitem [{\citenamefont {Cohen}\ \emph {et~al.}(2024)\citenamefont {Cohen}, \citenamefont {Samuelson}, \citenamefont {Wang}, \citenamefont {Klocke}, \citenamefont {Reeves}, \citenamefont {Taniguchi}, \citenamefont {Watanabe}, \citenamefont {Vijay}, \citenamefont {Zaletel},\ and\ \citenamefont {Young}}]{Young2024UCvdWLGG}%
  \BibitemOpen
  \bibfield  {author} {\bibinfo {author} {\bibfnamefont {Liam~A.}\ \bibnamefont {Cohen}}, \bibinfo {author} {\bibfnamefont {Noah~L.}\ \bibnamefont {Samuelson}}, \bibinfo {author} {\bibfnamefont {Taige}\ \bibnamefont {Wang}}, \bibinfo {author} {\bibfnamefont {Kai}\ \bibnamefont {Klocke}}, \bibinfo {author} {\bibfnamefont {Cian~C.}\ \bibnamefont {Reeves}}, \bibinfo {author} {\bibfnamefont {Takashi}\ \bibnamefont {Taniguchi}}, \bibinfo {author} {\bibfnamefont {Kenji}\ \bibnamefont {Watanabe}}, \bibinfo {author} {\bibfnamefont {Sagar}\ \bibnamefont {Vijay}}, \bibinfo {author} {\bibfnamefont {Michael~P.}\ \bibnamefont {Zaletel}}, \ and\ \bibinfo {author} {\bibfnamefont {Andrea~F.}\ \bibnamefont {Young}},\ }\bibfield  {title} {\enquote {\bibinfo {title} {Nanoscale electrostatic control in ultra clean van der waals heterostructures by local anodic oxidation of graphite gates},}\ }\href@noop {} {\bibfield  {journal} {\bibinfo  {journal} {arXiv preprint arXiv:2204.10296}\ } (\bibinfo {year} {2024})}\BibitemShut
  {NoStop}%
\bibitem [{\citenamefont {Ong}\ and\ \citenamefont {Fischetti}(2012)}]{ong2012impurity}%
  \BibitemOpen
  \bibfield  {author} {\bibinfo {author} {\bibfnamefont {Zhun-Yong}\ \bibnamefont {Ong}}\ and\ \bibinfo {author} {\bibfnamefont {Massimo~V}\ \bibnamefont {Fischetti}},\ }\bibfield  {title} {\enquote {\bibinfo {title} {Charged impurity scattering in top-gated graphene nanostructures},}\ }\href@noop {} {\bibfield  {journal} {\bibinfo  {journal} {Physical Review B—Condensed Matter and Materials Physics}\ }\textbf {\bibinfo {volume} {86}},\ \bibinfo {pages} {121409} (\bibinfo {year} {2012})}\BibitemShut {NoStop}%
\bibitem [{\citenamefont {Li}\ \emph {et~al.}(2021{\natexlab{b}})\citenamefont {Li}, \citenamefont {Li}, \citenamefont {Naik}, \citenamefont {Xie}, \citenamefont {Li}, \citenamefont {Regan}, \citenamefont {Wang}, \citenamefont {Zhao}, \citenamefont {Yumigeta}, \citenamefont {Blei} \emph {et~al.}}]{Li2021STM}%
  \BibitemOpen
  \bibfield  {author} {\bibinfo {author} {\bibfnamefont {Hongyuan}\ \bibnamefont {Li}}, \bibinfo {author} {\bibfnamefont {Shaowei}\ \bibnamefont {Li}}, \bibinfo {author} {\bibfnamefont {Mit~H}\ \bibnamefont {Naik}}, \bibinfo {author} {\bibfnamefont {Jingxu}\ \bibnamefont {Xie}}, \bibinfo {author} {\bibfnamefont {Xinyu}\ \bibnamefont {Li}}, \bibinfo {author} {\bibfnamefont {Emma}\ \bibnamefont {Regan}}, \bibinfo {author} {\bibfnamefont {Danqing}\ \bibnamefont {Wang}}, \bibinfo {author} {\bibfnamefont {Wenyu}\ \bibnamefont {Zhao}}, \bibinfo {author} {\bibfnamefont {Kentaro}\ \bibnamefont {Yumigeta}}, \bibinfo {author} {\bibfnamefont {Mark}\ \bibnamefont {Blei}},  \emph {et~al.},\ }\bibfield  {title} {\enquote {\bibinfo {title} {Imaging local discharge cascades for correlated electrons in ws2/wse2 moir{\'e} superlattices},}\ }\href@noop {} {\bibfield  {journal} {\bibinfo  {journal} {Nature Physics}\ }\textbf {\bibinfo {volume} {17}},\ \bibinfo {pages} {1114--1119} (\bibinfo {year}
  {2021}{\natexlab{b}})}\BibitemShut {NoStop}%
\bibitem [{\citenamefont {Xie}\ \emph {et~al.}(2021)\citenamefont {Xie}, \citenamefont {Pierce}, \citenamefont {Park}, \citenamefont {Parker}, \citenamefont {Khalaf}, \citenamefont {Ledwith}, \citenamefont {Cao}, \citenamefont {Lee}, \citenamefont {Chen}, \citenamefont {Forrester} \emph {et~al.}}]{Xie2021SET}%
  \BibitemOpen
  \bibfield  {author} {\bibinfo {author} {\bibfnamefont {Yonglong}\ \bibnamefont {Xie}}, \bibinfo {author} {\bibfnamefont {Andrew~T}\ \bibnamefont {Pierce}}, \bibinfo {author} {\bibfnamefont {Jeong~Min}\ \bibnamefont {Park}}, \bibinfo {author} {\bibfnamefont {Daniel~E}\ \bibnamefont {Parker}}, \bibinfo {author} {\bibfnamefont {Eslam}\ \bibnamefont {Khalaf}}, \bibinfo {author} {\bibfnamefont {Patrick}\ \bibnamefont {Ledwith}}, \bibinfo {author} {\bibfnamefont {Yuan}\ \bibnamefont {Cao}}, \bibinfo {author} {\bibfnamefont {Seung~Hwan}\ \bibnamefont {Lee}}, \bibinfo {author} {\bibfnamefont {Shaowen}\ \bibnamefont {Chen}}, \bibinfo {author} {\bibfnamefont {Patrick~R}\ \bibnamefont {Forrester}},  \emph {et~al.},\ }\bibfield  {title} {\enquote {\bibinfo {title} {Fractional chern insulators in magic-angle twisted bilayer graphene},}\ }\href@noop {} {\bibfield  {journal} {\bibinfo  {journal} {Nature}\ }\textbf {\bibinfo {volume} {600}},\ \bibinfo {pages} {439--443} (\bibinfo {year} {2021})}\BibitemShut {NoStop}%
\bibitem [{\citenamefont {Eisenstein}\ \emph {et~al.}(1992)\citenamefont {Eisenstein}, \citenamefont {Pfeiffer},\ and\ \citenamefont {West}}]{eisenstein1992negative}%
  \BibitemOpen
  \bibfield  {author} {\bibinfo {author} {\bibfnamefont {JP}~\bibnamefont {Eisenstein}}, \bibinfo {author} {\bibfnamefont {LN}~\bibnamefont {Pfeiffer}}, \ and\ \bibinfo {author} {\bibfnamefont {KW}~\bibnamefont {West}},\ }\bibfield  {title} {\enquote {\bibinfo {title} {Negative compressibility of interacting two-dimensional electron and quasiparticle gases},}\ }\href@noop {} {\bibfield  {journal} {\bibinfo  {journal} {Physical review letters}\ }\textbf {\bibinfo {volume} {68}},\ \bibinfo {pages} {674} (\bibinfo {year} {1992})}\BibitemShut {NoStop}%
\bibitem [{\citenamefont {Li}\ \emph {et~al.}(2021{\natexlab{c}})\citenamefont {Li}, \citenamefont {Li}, \citenamefont {Regan}, \citenamefont {Wang}, \citenamefont {Zhao}, \citenamefont {Kahn}, \citenamefont {Yumigeta}, \citenamefont {Blei}, \citenamefont {Taniguchi}, \citenamefont {Watanabe} \emph {et~al.}}]{Li2021STM_dual}%
  \BibitemOpen
  \bibfield  {author} {\bibinfo {author} {\bibfnamefont {Hongyuan}\ \bibnamefont {Li}}, \bibinfo {author} {\bibfnamefont {Shaowei}\ \bibnamefont {Li}}, \bibinfo {author} {\bibfnamefont {Emma~C}\ \bibnamefont {Regan}}, \bibinfo {author} {\bibfnamefont {Danqing}\ \bibnamefont {Wang}}, \bibinfo {author} {\bibfnamefont {Wenyu}\ \bibnamefont {Zhao}}, \bibinfo {author} {\bibfnamefont {Salman}\ \bibnamefont {Kahn}}, \bibinfo {author} {\bibfnamefont {Kentaro}\ \bibnamefont {Yumigeta}}, \bibinfo {author} {\bibfnamefont {Mark}\ \bibnamefont {Blei}}, \bibinfo {author} {\bibfnamefont {Takashi}\ \bibnamefont {Taniguchi}}, \bibinfo {author} {\bibfnamefont {Kenji}\ \bibnamefont {Watanabe}},  \emph {et~al.},\ }\bibfield  {title} {\enquote {\bibinfo {title} {Imaging two-dimensional generalized wigner crystals},}\ }\href@noop {} {\bibfield  {journal} {\bibinfo  {journal} {Nature}\ }\textbf {\bibinfo {volume} {597}},\ \bibinfo {pages} {650--654} (\bibinfo {year} {2021}{\natexlab{c}})}\BibitemShut {NoStop}%
\bibitem [{\citenamefont {Chiu}\ \emph {et~al.}(2024)\citenamefont {Chiu}, \citenamefont {Wang}, \citenamefont {Fan}, \citenamefont {Watanabe}, \citenamefont {Taniguchi}, \citenamefont {Liu}, \citenamefont {Zaletel},\ and\ \citenamefont {Yazdani}}]{Chiu2024STM_dual}%
  \BibitemOpen
  \bibfield  {author} {\bibinfo {author} {\bibfnamefont {Cheng-Li}\ \bibnamefont {Chiu}}, \bibinfo {author} {\bibfnamefont {Taige}\ \bibnamefont {Wang}}, \bibinfo {author} {\bibfnamefont {Ruihua}\ \bibnamefont {Fan}}, \bibinfo {author} {\bibfnamefont {Kenji}\ \bibnamefont {Watanabe}}, \bibinfo {author} {\bibfnamefont {Takashi}\ \bibnamefont {Taniguchi}}, \bibinfo {author} {\bibfnamefont {Xiaomeng}\ \bibnamefont {Liu}}, \bibinfo {author} {\bibfnamefont {Michael~P}\ \bibnamefont {Zaletel}}, \ and\ \bibinfo {author} {\bibfnamefont {Ali}\ \bibnamefont {Yazdani}},\ }\bibfield  {title} {\enquote {\bibinfo {title} {High spatial resolution charge sensing of quantum hall states},}\ }\href@noop {} {\bibfield  {journal} {\bibinfo  {journal} {arXiv preprint arXiv:2410.10961}\ } (\bibinfo {year} {2024})}\BibitemShut {NoStop}%
\bibitem [{\citenamefont {Cao}\ \emph {et~al.}(2023)\citenamefont {Cao}, \citenamefont {Wu}, \citenamefont {Bhattacharyya}, \citenamefont {Zhang},\ and\ \citenamefont {Allen}}]{Cao2023mK}%
  \BibitemOpen
  \bibfield  {author} {\bibinfo {author} {\bibfnamefont {Leonard~Weihao}\ \bibnamefont {Cao}}, \bibinfo {author} {\bibfnamefont {Chen}\ \bibnamefont {Wu}}, \bibinfo {author} {\bibfnamefont {Rajarshi}\ \bibnamefont {Bhattacharyya}}, \bibinfo {author} {\bibfnamefont {Ruolun}\ \bibnamefont {Zhang}}, \ and\ \bibinfo {author} {\bibfnamefont {Monica~T.}\ \bibnamefont {Allen}},\ }\bibfield  {title} {\enquote {\bibinfo {title} {Millikelvin microwave impedance microscopy in a dry dilution refrigerator},}\ }\href {\doibase 10.1063/5.0159548} {\bibfield  {journal} {\bibinfo  {journal} {Review of Scientific Instruments}\ }\textbf {\bibinfo {volume} {94}},\ \bibinfo {pages} {093705} (\bibinfo {year} {2023})}\BibitemShut {NoStop}%
\bibitem [{\citenamefont {Wang}\ \emph {et~al.}(2023)\citenamefont {Wang}, \citenamefont {Wu}, \citenamefont {Mogi}, \citenamefont {Kawamura}, \citenamefont {Tokura}, \citenamefont {Shen}, \citenamefont {You},\ and\ \citenamefont {Allen}}]{Taige2023comsol}%
  \BibitemOpen
  \bibfield  {author} {\bibinfo {author} {\bibfnamefont {Taige}\ \bibnamefont {Wang}}, \bibinfo {author} {\bibfnamefont {Chen}\ \bibnamefont {Wu}}, \bibinfo {author} {\bibfnamefont {Masataka}\ \bibnamefont {Mogi}}, \bibinfo {author} {\bibfnamefont {Minoru}\ \bibnamefont {Kawamura}}, \bibinfo {author} {\bibfnamefont {Yoshinori}\ \bibnamefont {Tokura}}, \bibinfo {author} {\bibfnamefont {Zhi-Xun}\ \bibnamefont {Shen}}, \bibinfo {author} {\bibfnamefont {Yi-Zhuang}\ \bibnamefont {You}}, \ and\ \bibinfo {author} {\bibfnamefont {Monica~T}\ \bibnamefont {Allen}},\ }\bibfield  {title} {\enquote {\bibinfo {title} {Probing the edge states of chern insulators using microwave impedance microscopy},}\ }\href@noop {} {\bibfield  {journal} {\bibinfo  {journal} {Physical Review B}\ }\textbf {\bibinfo {volume} {108}},\ \bibinfo {pages} {235432} (\bibinfo {year} {2023})}\BibitemShut {NoStop}%
\bibitem [{\citenamefont {Castro~Neto}\ \emph {et~al.}(2009)\citenamefont {Castro~Neto}, \citenamefont {Guinea}, \citenamefont {Peres}, \citenamefont {Novoselov},\ and\ \citenamefont {Geim}}]{castro2009electronic}%
  \BibitemOpen
  \bibfield  {author} {\bibinfo {author} {\bibfnamefont {Antonio~H}\ \bibnamefont {Castro~Neto}}, \bibinfo {author} {\bibfnamefont {Francisco}\ \bibnamefont {Guinea}}, \bibinfo {author} {\bibfnamefont {Nuno~MR}\ \bibnamefont {Peres}}, \bibinfo {author} {\bibfnamefont {Kostya~S}\ \bibnamefont {Novoselov}}, \ and\ \bibinfo {author} {\bibfnamefont {Andre~K}\ \bibnamefont {Geim}},\ }\bibfield  {title} {\enquote {\bibinfo {title} {The electronic properties of graphene},}\ }\href@noop {} {\bibfield  {journal} {\bibinfo  {journal} {Reviews of modern physics}\ }\textbf {\bibinfo {volume} {81}},\ \bibinfo {pages} {109--162} (\bibinfo {year} {2009})}\BibitemShut {NoStop}%
\bibitem [{\citenamefont {Abanin}\ \emph {et~al.}(2013)\citenamefont {Abanin}, \citenamefont {Feldman}, \citenamefont {Yacoby},\ and\ \citenamefont {Halperin}}]{Halperin2013}%
  \BibitemOpen
  \bibfield  {author} {\bibinfo {author} {\bibfnamefont {Dmitry~A.}\ \bibnamefont {Abanin}}, \bibinfo {author} {\bibfnamefont {Benjamin~E.}\ \bibnamefont {Feldman}}, \bibinfo {author} {\bibfnamefont {Amir}\ \bibnamefont {Yacoby}}, \ and\ \bibinfo {author} {\bibfnamefont {Bertrand~I.}\ \bibnamefont {Halperin}},\ }\bibfield  {title} {\enquote {\bibinfo {title} {Fractional and integer quantum hall effects in the zeroth landau level in graphene},}\ }\href {\doibase 10.1103/PhysRevB.88.115407} {\bibfield  {journal} {\bibinfo  {journal} {Phys. Rev. B}\ }\textbf {\bibinfo {volume} {88}},\ \bibinfo {pages} {115407} (\bibinfo {year} {2013})}\BibitemShut {NoStop}%
\bibitem [{\citenamefont {Eisenstein}\ \emph {et~al.}(1994)\citenamefont {Eisenstein}, \citenamefont {Pfeiffer},\ and\ \citenamefont {West}}]{Eisenstein1994compressibility}%
  \BibitemOpen
  \bibfield  {author} {\bibinfo {author} {\bibfnamefont {J.~P.}\ \bibnamefont {Eisenstein}}, \bibinfo {author} {\bibfnamefont {L.~N.}\ \bibnamefont {Pfeiffer}}, \ and\ \bibinfo {author} {\bibfnamefont {K.~W.}\ \bibnamefont {West}},\ }\bibfield  {title} {\enquote {\bibinfo {title} {Compressibility of the two-dimensional electron gas: Measurements of the zero-field exchange energy and fractional quantum hall gap},}\ }\href {\doibase 10.1103/PhysRevB.50.1760} {\bibfield  {journal} {\bibinfo  {journal} {Phys. Rev. B}\ }\textbf {\bibinfo {volume} {50}},\ \bibinfo {pages} {1760--1778} (\bibinfo {year} {1994})}\BibitemShut {NoStop}%
\bibitem [{\citenamefont {Kim}\ \emph {et~al.}(2012)\citenamefont {Kim}, \citenamefont {Jo}, \citenamefont {Dillen}, \citenamefont {Ferrer}, \citenamefont {Fallahazad}, \citenamefont {Yao}, \citenamefont {Banerjee},\ and\ \citenamefont {Tutuc}}]{Tutuc2012Fermi}%
  \BibitemOpen
  \bibfield  {author} {\bibinfo {author} {\bibfnamefont {Seyoung}\ \bibnamefont {Kim}}, \bibinfo {author} {\bibfnamefont {Insun}\ \bibnamefont {Jo}}, \bibinfo {author} {\bibfnamefont {D.~C.}\ \bibnamefont {Dillen}}, \bibinfo {author} {\bibfnamefont {D.~A.}\ \bibnamefont {Ferrer}}, \bibinfo {author} {\bibfnamefont {B.}~\bibnamefont {Fallahazad}}, \bibinfo {author} {\bibfnamefont {Z.}~\bibnamefont {Yao}}, \bibinfo {author} {\bibfnamefont {S.~K.}\ \bibnamefont {Banerjee}}, \ and\ \bibinfo {author} {\bibfnamefont {E.}~\bibnamefont {Tutuc}},\ }\bibfield  {title} {\enquote {\bibinfo {title} {Direct measurement of the fermi energy in graphene using a double-layer heterostructure},}\ }\href {\doibase 10.1103/PhysRevLett.108.116404} {\bibfield  {journal} {\bibinfo  {journal} {Phys. Rev. Lett.}\ }\textbf {\bibinfo {volume} {108}},\ \bibinfo {pages} {116404} (\bibinfo {year} {2012})}\BibitemShut {NoStop}%
\bibitem [{\citenamefont {Kim}\ \emph {et~al.}(2021)\citenamefont {Kim}, \citenamefont {Schwenk}, \citenamefont {Walkup}, \citenamefont {Zeng}, \citenamefont {Ghahari}, \citenamefont {Le}, \citenamefont {Slot}, \citenamefont {Berwanger}, \citenamefont {Blankenship}, \citenamefont {Watanabe} \emph {et~al.}}]{kim2021AFM}%
  \BibitemOpen
  \bibfield  {author} {\bibinfo {author} {\bibfnamefont {Sungmin}\ \bibnamefont {Kim}}, \bibinfo {author} {\bibfnamefont {Johannes}\ \bibnamefont {Schwenk}}, \bibinfo {author} {\bibfnamefont {Daniel}\ \bibnamefont {Walkup}}, \bibinfo {author} {\bibfnamefont {Yihang}\ \bibnamefont {Zeng}}, \bibinfo {author} {\bibfnamefont {Fereshte}\ \bibnamefont {Ghahari}}, \bibinfo {author} {\bibfnamefont {Son~T}\ \bibnamefont {Le}}, \bibinfo {author} {\bibfnamefont {Marlou~R}\ \bibnamefont {Slot}}, \bibinfo {author} {\bibfnamefont {Julian}\ \bibnamefont {Berwanger}}, \bibinfo {author} {\bibfnamefont {Steven~R}\ \bibnamefont {Blankenship}}, \bibinfo {author} {\bibfnamefont {Kenji}\ \bibnamefont {Watanabe}},  \emph {et~al.},\ }\bibfield  {title} {\enquote {\bibinfo {title} {Edge channels of broken-symmetry quantum hall states in graphene visualized by atomic force microscopy},}\ }\href@noop {} {\bibfield  {journal} {\bibinfo  {journal} {Nature Communications}\ }\textbf {\bibinfo {volume} {12}},\ \bibinfo {pages} {2852}
  (\bibinfo {year} {2021})}\BibitemShut {NoStop}%
\bibitem [{\citenamefont {Novoselov}\ \emph {et~al.}(2005)\citenamefont {Novoselov}, \citenamefont {Geim}, \citenamefont {Morozov}, \citenamefont {Jiang}, \citenamefont {Katsnelson}, \citenamefont {Grigorieva}, \citenamefont {Dubonos},\ and\ \citenamefont {Firsov}}]{novoselov2005two}%
  \BibitemOpen
  \bibfield  {author} {\bibinfo {author} {\bibfnamefont {Kostya~S}\ \bibnamefont {Novoselov}}, \bibinfo {author} {\bibfnamefont {Andre~K}\ \bibnamefont {Geim}}, \bibinfo {author} {\bibfnamefont {Sergei~Vladimirovich}\ \bibnamefont {Morozov}}, \bibinfo {author} {\bibfnamefont {Dingde}\ \bibnamefont {Jiang}}, \bibinfo {author} {\bibfnamefont {Michail~I}\ \bibnamefont {Katsnelson}}, \bibinfo {author} {\bibfnamefont {Irina~V}\ \bibnamefont {Grigorieva}}, \bibinfo {author} {\bibfnamefont {Sergey~V}\ \bibnamefont {Dubonos}}, \ and\ \bibinfo {author} {\bibfnamefont {Alexandr~A}\ \bibnamefont {Firsov}},\ }\bibfield  {title} {\enquote {\bibinfo {title} {Two-dimensional gas of massless dirac fermions in graphene},}\ }\href@noop {} {\bibfield  {journal} {\bibinfo  {journal} {nature}\ }\textbf {\bibinfo {volume} {438}},\ \bibinfo {pages} {197--200} (\bibinfo {year} {2005})}\BibitemShut {NoStop}%
\bibitem [{\citenamefont {Park}\ \emph {et~al.}(2020)\citenamefont {Park}, \citenamefont {Cao}, \citenamefont {Watanabe}, \citenamefont {Taniguchi},\ and\ \citenamefont {Jarillo-Herrero}}]{park2020flavour}%
  \BibitemOpen
  \bibfield  {author} {\bibinfo {author} {\bibfnamefont {Jeong~Min}\ \bibnamefont {Park}}, \bibinfo {author} {\bibfnamefont {Yuan}\ \bibnamefont {Cao}}, \bibinfo {author} {\bibfnamefont {Kenji}\ \bibnamefont {Watanabe}}, \bibinfo {author} {\bibfnamefont {Takashi}\ \bibnamefont {Taniguchi}}, \ and\ \bibinfo {author} {\bibfnamefont {Pablo}\ \bibnamefont {Jarillo-Herrero}},\ }\bibfield  {title} {\enquote {\bibinfo {title} {Flavour hund's coupling, correlated chern gaps, and diffusivity in moir$\backslash$'e flat bands},}\ }\href@noop {} {\bibfield  {journal} {\bibinfo  {journal} {arXiv preprint arXiv:2008.12296}\ } (\bibinfo {year} {2020})}\BibitemShut {NoStop}%
\bibitem [{\citenamefont {Zondiner}\ \emph {et~al.}(2020)\citenamefont {Zondiner}, \citenamefont {Rozen}, \citenamefont {Rodan-Legrain}, \citenamefont {Cao}, \citenamefont {Queiroz}, \citenamefont {Taniguchi}, \citenamefont {Watanabe}, \citenamefont {Oreg}, \citenamefont {von Oppen}, \citenamefont {Stern} \emph {et~al.}}]{zondiner2020cascade}%
  \BibitemOpen
  \bibfield  {author} {\bibinfo {author} {\bibfnamefont {Uri}\ \bibnamefont {Zondiner}}, \bibinfo {author} {\bibfnamefont {Asaf}\ \bibnamefont {Rozen}}, \bibinfo {author} {\bibfnamefont {Daniel}\ \bibnamefont {Rodan-Legrain}}, \bibinfo {author} {\bibfnamefont {Yuan}\ \bibnamefont {Cao}}, \bibinfo {author} {\bibfnamefont {Raquel}\ \bibnamefont {Queiroz}}, \bibinfo {author} {\bibfnamefont {Takashi}\ \bibnamefont {Taniguchi}}, \bibinfo {author} {\bibfnamefont {Kenji}\ \bibnamefont {Watanabe}}, \bibinfo {author} {\bibfnamefont {Yuval}\ \bibnamefont {Oreg}}, \bibinfo {author} {\bibfnamefont {Felix}\ \bibnamefont {von Oppen}}, \bibinfo {author} {\bibfnamefont {Ady}\ \bibnamefont {Stern}},  \emph {et~al.},\ }\bibfield  {title} {\enquote {\bibinfo {title} {Cascade of phase transitions and dirac revivals in magic-angle graphene},}\ }\href@noop {} {\bibfield  {journal} {\bibinfo  {journal} {Nature}\ }\textbf {\bibinfo {volume} {582}},\ \bibinfo {pages} {203--208} (\bibinfo {year} {2020})}\BibitemShut {NoStop}%
\bibitem [{\citenamefont {Li}\ \emph {et~al.}(2009)\citenamefont {Li}, \citenamefont {Vicente}, \citenamefont {Xia}, \citenamefont {Pan}, \citenamefont {Tsui}, \citenamefont {Pfeiffer},\ and\ \citenamefont {West}}]{PhysRevLett.102.216801}%
  \BibitemOpen
  \bibfield  {author} {\bibinfo {author} {\bibfnamefont {Wanli}\ \bibnamefont {Li}}, \bibinfo {author} {\bibfnamefont {C.~L.}\ \bibnamefont {Vicente}}, \bibinfo {author} {\bibfnamefont {J.~S.}\ \bibnamefont {Xia}}, \bibinfo {author} {\bibfnamefont {W.}~\bibnamefont {Pan}}, \bibinfo {author} {\bibfnamefont {D.~C.}\ \bibnamefont {Tsui}}, \bibinfo {author} {\bibfnamefont {L.~N.}\ \bibnamefont {Pfeiffer}}, \ and\ \bibinfo {author} {\bibfnamefont {K.~W.}\ \bibnamefont {West}},\ }\bibfield  {title} {\enquote {\bibinfo {title} {Scaling in plateau-to-plateau transition: A direct connection of quantum hall systems with the anderson localization model},}\ }\href {\doibase 10.1103/PhysRevLett.102.216801} {\bibfield  {journal} {\bibinfo  {journal} {Phys. Rev. Lett.}\ }\textbf {\bibinfo {volume} {102}},\ \bibinfo {pages} {216801} (\bibinfo {year} {2009})}\BibitemShut {NoStop}%
\bibitem [{\citenamefont {Arovas}()}]{arovasFQHEnotes}%
  \BibitemOpen
  \bibfield  {author} {\bibinfo {author} {\bibfnamefont {Dan}\ \bibnamefont {Arovas}},\ }\href@noop {} {\enquote {\bibinfo {title} {Lecture notes on quantum hall effect},}\ }\bibinfo {howpublished} {\url{https://courses.physics.ucsd.edu/2019/Spring/physics230/LECTURES/QHE.pdf}},\ \bibinfo {note} {accessed: 2025-6-24}\BibitemShut {NoStop}%
\bibitem [{\citenamefont {Puschmann}\ \emph {et~al.}(2019)\citenamefont {Puschmann}, \citenamefont {Cain}, \citenamefont {Schreiber},\ and\ \citenamefont {Vojta}}]{PhysRevB.99.121301}%
  \BibitemOpen
  \bibfield  {author} {\bibinfo {author} {\bibfnamefont {Martin}\ \bibnamefont {Puschmann}}, \bibinfo {author} {\bibfnamefont {Philipp}\ \bibnamefont {Cain}}, \bibinfo {author} {\bibfnamefont {Michael}\ \bibnamefont {Schreiber}}, \ and\ \bibinfo {author} {\bibfnamefont {Thomas}\ \bibnamefont {Vojta}},\ }\bibfield  {title} {\enquote {\bibinfo {title} {Integer quantum hall transition on a tight-binding lattice},}\ }\href {\doibase 10.1103/PhysRevB.99.121301} {\bibfield  {journal} {\bibinfo  {journal} {Phys. Rev. B}\ }\textbf {\bibinfo {volume} {99}},\ \bibinfo {pages} {121301} (\bibinfo {year} {2019})}\BibitemShut {NoStop}%
\bibitem [{\citenamefont {Chalker}\ and\ \citenamefont {Coddington}(1988)}]{chalker1988percolation}%
  \BibitemOpen
  \bibfield  {author} {\bibinfo {author} {\bibfnamefont {JT}~\bibnamefont {Chalker}}\ and\ \bibinfo {author} {\bibfnamefont {PD}~\bibnamefont {Coddington}},\ }\bibfield  {title} {\enquote {\bibinfo {title} {Percolation, quantum tunnelling and the integer hall effect},}\ }\href@noop {} {\bibfield  {journal} {\bibinfo  {journal} {Journal of Physics C: Solid State Physics}\ }\textbf {\bibinfo {volume} {21}},\ \bibinfo {pages} {2665} (\bibinfo {year} {1988})}\BibitemShut {NoStop}%
\bibitem [{\citenamefont {Li}\ \emph {et~al.}(2013)\citenamefont {Li}, \citenamefont {Luican-Mayer}, \citenamefont {Abanin}, \citenamefont {Levitov},\ and\ \citenamefont {Andrei}}]{li2013evolution}%
  \BibitemOpen
  \bibfield  {author} {\bibinfo {author} {\bibfnamefont {Guohong}\ \bibnamefont {Li}}, \bibinfo {author} {\bibfnamefont {Adina}\ \bibnamefont {Luican-Mayer}}, \bibinfo {author} {\bibfnamefont {Dmitry}\ \bibnamefont {Abanin}}, \bibinfo {author} {\bibfnamefont {Leonid}\ \bibnamefont {Levitov}}, \ and\ \bibinfo {author} {\bibfnamefont {Eva~Y}\ \bibnamefont {Andrei}},\ }\bibfield  {title} {\enquote {\bibinfo {title} {Evolution of landau levels into edge states in graphene},}\ }\href@noop {} {\bibfield  {journal} {\bibinfo  {journal} {Nature communications}\ }\textbf {\bibinfo {volume} {4}},\ \bibinfo {pages} {1744} (\bibinfo {year} {2013})}\BibitemShut {NoStop}%
\bibitem [{\citenamefont {Coissard}\ \emph {et~al.}(2023)\citenamefont {Coissard}, \citenamefont {Grushin}, \citenamefont {Repellin}, \citenamefont {Veyrat}, \citenamefont {Watanabe}, \citenamefont {Taniguchi}, \citenamefont {Gay}, \citenamefont {Courtois}, \citenamefont {Sellier},\ and\ \citenamefont {Sac{\'e}p{\'e}}}]{coissard2023absence}%
  \BibitemOpen
  \bibfield  {author} {\bibinfo {author} {\bibfnamefont {Alexis}\ \bibnamefont {Coissard}}, \bibinfo {author} {\bibfnamefont {Adolfo~G}\ \bibnamefont {Grushin}}, \bibinfo {author} {\bibfnamefont {C{\'e}cile}\ \bibnamefont {Repellin}}, \bibinfo {author} {\bibfnamefont {Louis}\ \bibnamefont {Veyrat}}, \bibinfo {author} {\bibfnamefont {Kenji}\ \bibnamefont {Watanabe}}, \bibinfo {author} {\bibfnamefont {Takashi}\ \bibnamefont {Taniguchi}}, \bibinfo {author} {\bibfnamefont {Fr{\'e}d{\'e}ric}\ \bibnamefont {Gay}}, \bibinfo {author} {\bibfnamefont {Herv{\'e}}\ \bibnamefont {Courtois}}, \bibinfo {author} {\bibfnamefont {Hermann}\ \bibnamefont {Sellier}}, \ and\ \bibinfo {author} {\bibfnamefont {Benjamin}\ \bibnamefont {Sac{\'e}p{\'e}}},\ }\bibfield  {title} {\enquote {\bibinfo {title} {Absence of edge reconstruction for quantum hall edge channels in graphene devices},}\ }\href@noop {} {\bibfield  {journal} {\bibinfo  {journal} {Science Advances}\ }\textbf {\bibinfo {volume} {9}},\ \bibinfo {pages} {eadf7220}
  (\bibinfo {year} {2023})}\BibitemShut {NoStop}%
\bibitem [{\citenamefont {Liu}\ \emph {et~al.}(2022)\citenamefont {Liu}, \citenamefont {Farahi}, \citenamefont {Chiu}, \citenamefont {Papic}, \citenamefont {Watanabe}, \citenamefont {Taniguchi}, \citenamefont {Zaletel},\ and\ \citenamefont {Yazdani}}]{liu2022visualizing}%
  \BibitemOpen
  \bibfield  {author} {\bibinfo {author} {\bibfnamefont {Xiaomeng}\ \bibnamefont {Liu}}, \bibinfo {author} {\bibfnamefont {Gelareh}\ \bibnamefont {Farahi}}, \bibinfo {author} {\bibfnamefont {Cheng-Li}\ \bibnamefont {Chiu}}, \bibinfo {author} {\bibfnamefont {Zlatko}\ \bibnamefont {Papic}}, \bibinfo {author} {\bibfnamefont {Kenji}\ \bibnamefont {Watanabe}}, \bibinfo {author} {\bibfnamefont {Takashi}\ \bibnamefont {Taniguchi}}, \bibinfo {author} {\bibfnamefont {Michael~P}\ \bibnamefont {Zaletel}}, \ and\ \bibinfo {author} {\bibfnamefont {Ali}\ \bibnamefont {Yazdani}},\ }\bibfield  {title} {\enquote {\bibinfo {title} {Visualizing broken symmetry and topological defects in a quantum hall ferromagnet},}\ }\href@noop {} {\bibfield  {journal} {\bibinfo  {journal} {Science}\ }\textbf {\bibinfo {volume} {375}},\ \bibinfo {pages} {321--326} (\bibinfo {year} {2022})}\BibitemShut {NoStop}%
\bibitem [{\citenamefont {Rashidi}\ \emph {et~al.}(2023)\citenamefont {Rashidi}, \citenamefont {Kealhofer}, \citenamefont {Lygo},\ and\ \citenamefont {Stemmer}}]{rashidi2023Cd3As2}%
  \BibitemOpen
  \bibfield  {author} {\bibinfo {author} {\bibfnamefont {Arman}\ \bibnamefont {Rashidi}}, \bibinfo {author} {\bibfnamefont {Robert}\ \bibnamefont {Kealhofer}}, \bibinfo {author} {\bibfnamefont {Alexander~C}\ \bibnamefont {Lygo}}, \ and\ \bibinfo {author} {\bibfnamefont {Susanne}\ \bibnamefont {Stemmer}},\ }\bibfield  {title} {\enquote {\bibinfo {title} {Universal conductance fluctuations in nanoscale topological insulator devices},}\ }\href@noop {} {\bibfield  {journal} {\bibinfo  {journal} {Applied Physics Letters}\ }\textbf {\bibinfo {volume} {122}} (\bibinfo {year} {2023})}\BibitemShut {NoStop}%
\bibitem [{\citenamefont {Munyan}\ \emph {et~al.}(2023)\citenamefont {Munyan}, \citenamefont {Guo}, \citenamefont {Huynh}, \citenamefont {Huang},\ and\ \citenamefont {Stemmer}}]{munyan2023Cd3As2}%
  \BibitemOpen
  \bibfield  {author} {\bibinfo {author} {\bibfnamefont {Simon}\ \bibnamefont {Munyan}}, \bibinfo {author} {\bibfnamefont {Binghao}\ \bibnamefont {Guo}}, \bibinfo {author} {\bibfnamefont {William}\ \bibnamefont {Huynh}}, \bibinfo {author} {\bibfnamefont {Victor}\ \bibnamefont {Huang}}, \ and\ \bibinfo {author} {\bibfnamefont {Susanne}\ \bibnamefont {Stemmer}},\ }\bibfield  {title} {\enquote {\bibinfo {title} {Edge mode percolation and equilibration in the topological insulator cadmium arsenide},}\ }\href@noop {} {\bibfield  {journal} {\bibinfo  {journal} {npj Quantum Materials}\ }\textbf {\bibinfo {volume} {8}},\ \bibinfo {pages} {70} (\bibinfo {year} {2023})}\BibitemShut {NoStop}%
\bibitem [{\citenamefont {Mogi}\ \emph {et~al.}(2017)\citenamefont {Mogi}, \citenamefont {Kawamura}, \citenamefont {Yoshimi}, \citenamefont {Tsukazaki}, \citenamefont {Kozuka}, \citenamefont {Shirakawa}, \citenamefont {Takahashi}, \citenamefont {Kawasaki},\ and\ \citenamefont {Tokura}}]{mogi2017BiSbTe}%
  \BibitemOpen
  \bibfield  {author} {\bibinfo {author} {\bibfnamefont {M}~\bibnamefont {Mogi}}, \bibinfo {author} {\bibfnamefont {M}~\bibnamefont {Kawamura}}, \bibinfo {author} {\bibfnamefont {R}~\bibnamefont {Yoshimi}}, \bibinfo {author} {\bibfnamefont {A}~\bibnamefont {Tsukazaki}}, \bibinfo {author} {\bibfnamefont {Y}~\bibnamefont {Kozuka}}, \bibinfo {author} {\bibfnamefont {N}~\bibnamefont {Shirakawa}}, \bibinfo {author} {\bibfnamefont {KS}~\bibnamefont {Takahashi}}, \bibinfo {author} {\bibfnamefont {M}~\bibnamefont {Kawasaki}}, \ and\ \bibinfo {author} {\bibfnamefont {Y}~\bibnamefont {Tokura}},\ }\bibfield  {title} {\enquote {\bibinfo {title} {A magnetic heterostructure of topological insulators as a candidate for an axion insulator},}\ }\href@noop {} {\bibfield  {journal} {\bibinfo  {journal} {Nature materials}\ }\textbf {\bibinfo {volume} {16}},\ \bibinfo {pages} {516--521} (\bibinfo {year} {2017})}\BibitemShut {NoStop}%
\bibitem [{\citenamefont {Goossens}\ \emph {et~al.}(2012{\natexlab{b}})\citenamefont {Goossens}, \citenamefont {Calado}, \citenamefont {Barreiro}, \citenamefont {Watanabe}, \citenamefont {Taniguchi},\ and\ \citenamefont {Vandersypen}}]{goossens2012AFMClean}%
  \BibitemOpen
  \bibfield  {author} {\bibinfo {author} {\bibfnamefont {AM}~\bibnamefont {Goossens}}, \bibinfo {author} {\bibfnamefont {VE}~\bibnamefont {Calado}}, \bibinfo {author} {\bibfnamefont {A}~\bibnamefont {Barreiro}}, \bibinfo {author} {\bibfnamefont {K}~\bibnamefont {Watanabe}}, \bibinfo {author} {\bibfnamefont {T}~\bibnamefont {Taniguchi}}, \ and\ \bibinfo {author} {\bibfnamefont {LMK}\ \bibnamefont {Vandersypen}},\ }\bibfield  {title} {\enquote {\bibinfo {title} {Mechanical cleaning of graphene},}\ }\href@noop {} {\bibfield  {journal} {\bibinfo  {journal} {Applied Physics Letters}\ }\textbf {\bibinfo {volume} {100}} (\bibinfo {year} {2012}{\natexlab{b}})}\BibitemShut {NoStop}%
\bibitem [{\citenamefont {Cui}\ \emph {et~al.}(2016{\natexlab{b}})\citenamefont {Cui}, \citenamefont {Ma},\ and\ \citenamefont {Shen}}]{Cui2016tf}%
  \BibitemOpen
  \bibfield  {author} {\bibinfo {author} {\bibfnamefont {Yong-Tao}\ \bibnamefont {Cui}}, \bibinfo {author} {\bibfnamefont {Eric~Yue}\ \bibnamefont {Ma}}, \ and\ \bibinfo {author} {\bibfnamefont {Zhi-Xun}\ \bibnamefont {Shen}},\ }\bibfield  {title} {\enquote {\bibinfo {title} {Quartz tuning fork based microwave impedance microscopy},}\ }\href@noop {} {\bibfield  {journal} {\bibinfo  {journal} {Review of Scientific Instruments}\ }\textbf {\bibinfo {volume} {87}} (\bibinfo {year} {2016}{\natexlab{b}})}\BibitemShut {NoStop}%
\bibitem [{\citenamefont {Barber}\ \emph {et~al.}(2022)\citenamefont {Barber}, \citenamefont {Ma},\ and\ \citenamefont {Shen}}]{Barber2022MIM}%
  \BibitemOpen
  \bibfield  {author} {\bibinfo {author} {\bibfnamefont {Mark~E}\ \bibnamefont {Barber}}, \bibinfo {author} {\bibfnamefont {Eric~Yue}\ \bibnamefont {Ma}}, \ and\ \bibinfo {author} {\bibfnamefont {Zhi-Xun}\ \bibnamefont {Shen}},\ }\bibfield  {title} {\enquote {\bibinfo {title} {Microwave impedance microscopy and its application to quantum materials},}\ }\href@noop {} {\bibfield  {journal} {\bibinfo  {journal} {Nature Reviews Physics}\ }\textbf {\bibinfo {volume} {4}},\ \bibinfo {pages} {61--74} (\bibinfo {year} {2022})}\BibitemShut {NoStop}%
\end{thebibliography}%

\begin{appendix}

\renewcommand{\theequation}{S\arabic{equation}}
\setcounter{equation}{0}

\section*{Supplementary Information}

\textbf{Experimental methods.} The dual-gated graphene device was fabricated using a standard dry transfer method. Both hBN and graphite were mechanically exfoliated onto silicon (Si)/silicon dioxide (SiO$_2$) substrates, and suitable flakes were identified using an optical microscope. Graphene and graphite flakes went through thermal annealing in a H$_2$/N$_2$ condition at 350$^{\circ}$C for 6 hours. A PDMS stamp covered with a thin polycarbonate film was used to pick up flakes in the following order (from top to bottom): hBN - graphene - hBN - graphene - hBN - graphite. The entire heterostructure was deposited onto a substrate with pre-patterned contacts (Ti/Pt with thickness of 2nm/10nm) at 180$^{\circ}$C. After dissolving the polycarbonate film in chloroform, mechanical AFM cleaning was performed on the surface of the heterostructure \cite{goossens2012AFMClean}.

On the instrument side, the microwave probe in our setup is a thin (\SI{25}{\mu m} diameter) tungsten wire that is chemically etched to have a tapered-shape apex with a diameter around \SI{100}{nm} \cite{Cui2016tf,Cao2023mK}. An SEM image of the tip is shown in the inset of Fig.~\ref{fig:Schematics}(a). The tip is glued to one prong of the tuning fork and electrically soldered to the impedance matching network (IMN) in the cryogenic RF electronics \cite{Cao2023mK}. During the measurements, the tip apex is scanned above the sample surface with a tip-sample distance \SI{65}{nm}. In this experiment, a \SI{-23}{dBm} microwave excitation at \SI{4.22}{GHz} frequency is applied to the tip apex. The MIM signal reflected from the tip passes through the IMN to minimize signal loss from impedance mismatch, then travels to two directional couplers to cancel out the residual reflected power due to the mismatch, and finally is amplified by the cryogenic amplifier at \SI{4}{K} to improve the signal to noise ratio\cite{Cui2016tf,Cao2023mK}. At room temperature, the signal is further amplified and demodulated into the in-phase and out-of-phase MIM signal (raw MIM-Re and MIM-Im) with respect to the reference \cite{Barber2022MIM,Cao2023mK}. Finally, we use lock-in amplifiers to lock into the differential MIM signals embedded in the raw one, which is the height-modulated MIM signal generated by the tuning fork oscillation $d( \mathrm{MIM})/dz$ \cite{Cui2016tf,Cao2023mK}. All the MIM data presented in the article is measured with the height-modulated MIM.\\

\textbf{Band diagram models.}
When the top graphene layer is not in the filling $\nu_{\text{T}} = 0$, extra charge and chemical potential shift in the top graphene layer is introduced in Eq.~\eqref{eq:voltage_carrier_relations}. Since the function of chemical potential versus carrier density is monotonously non-decreasing for graphene in the single particle picture,  the deviation from the $n_\text{T} = 0$ state gives   $e\Delta V_{\text{TG}} = -\mu_{\text{T}}(n_{\text{T}}) - e^2 \Delta n_{\text{T}}/C_\text{M}$ and $
e \Delta V_{\text{BG}} = e^2 \Delta n_{\text{T}}/C_\text{B} $, the different top layer fillings $\nu_{\text{T}}$ are anti-diagonally spaced in the $V_{\text{TG}}$ - $V_{\text{BG}}$ plane. An illustration of the dependence of top graphene filling factors and the two gate voltages can be found in Fig.~\ref{fig:LL}(a).

To understand the expected MIM response of this system, we employ finite-element analysis (FEA) to solve the electrodynamics at RF frequencies in the linear-response regime. The MIM tip-sample interaction is modeled by approximating the MIM tip as a conical shape with a slightly rounded tip apex, which more accurately reflects the tip geometry in the experiment than a simplified point-charge model. The electrical properties of tungsten are incorporated into the model. The sample heterostructure is modeled as two thin conductive layers with well-defined longitudinal conductivity, representing the top graphene layer and the bottom graphene layer, separated by a dielectric layer of hBN. To compute the longitudinal conductivity, $\sigma^{xx}$, we first calculate the carrier density for each layer under a given combination of $V_{\text{TG}}$ and $V_{\text{BG}}$ according to equations (\ref{eqn:Tutuc1} and \ref{eqn:Tutuc2}). At a finite magnetic field $B$, the broadening of the density of states (DOS), $g(E)$, needs to be considered as \cite{Tutuc2012Fermi}:
\begin{equation}\label{eqn:DOE}
g(E) = \frac{4e}{h} B\sum_N\frac{1}{\pi}\frac{\gamma_N}{(E-E_N)^2 + \gamma_N^2}
\end{equation}
where $\gamma_N$ is broadening factor of the $N^{\textbf{th}}$ Landau level. The carrier densities $n_\mathrm{B}$ and $n_\mathrm{T}$ in equations \ref{eqn:Tutuc1} and \ref{eqn:Tutuc2} are calculated by integrating the DOS $g(E)$ up to the chemical potentials $\mu_\mathrm{B}$ and $\mu_\mathrm{T}$, respectively. With calculated $n_{\text{B}}$ and $n_{\text{T}}$ for the individual layers, we extract the conductivity tensor at a finite magnetic field $B$ based on the experimental data from Ref.~\cite{novoselov2005two}. The longitudinal conductivities of the top layer $\sigma^{xx}_\mathrm{T}$ and the bottom layer $\sigma^{xx}_\mathrm{B}$ are shown as color plots in Fig.~\ref{fig:LL}(c-d), as a function of $V_{\text{TG}}$ and $V_{\text{BG}}$. In  Fig.~\ref{fig:LL}(d), we define the perpendicular displacement field, $D$, on the bottom graphene layer using the formula $ D / \epsilon_0 = (C_{\text{B}}V_{\text{BG}} - 
 C_{\text{M}} V_{\text{TG}}) / 2\epsilon_0$.\\

\textbf{Fermi energy extraction.} Here we describe the procedure for extracting the chemical potential $\mu_{\text{B}}$, carrier density $n_{\text{B}}$, and other physical quantities of the sample layer from the MIM data. We begin by selecting a specific filling factor in the top graphene sensor layer ($\nu_{\text{T}}$) and tracing its trajectory in the $V_{\text{BG}} - V_{\text{TG}}$ plane. These trajectories, marked by pink dotted lines in Fig.~\ref{fig:LL}(a), follow the minima in the MIM-Im signal, which correspond to highly incompressible or gapped quantum Hall states in the sensor layer.  

At each data point along the trajectory corresponding to a fixed $\nu_{\text{T}}$, we can plug the corresponding values of $V_{\text{TG}}$ and $V_{\text{BG}}$ into Eq. \eqref{eqn:Tutuc1} and \eqref{eqn:Tutuc2}. At a given magnetic field $B$, it is straightforward to calculate the carrier density $n_{\text{T}}$ and the chemical potential $\mu_{\text{T}}$ of the top ``sensor'' layer based on the filling factor $\nu_{\text{T}}$. With this information,  $n_{\text{B}}$ and $\mu_{\text{B}}$ can be solved for the bottom (sample) layer. To reduce fluctuations arising from noise sources such as pulse tube vibrations, we apply a horizontal moving average over three adjacent data points along each converted line cut.
\\

\textbf{{Linear response model of MIM}}. 
MIM is a probe of the charge susceptibility $\chi$ (i.e. the dynamical density-density correlator) of the sample. In the case where the tip is suspended a distance $d$ above the sample and separated by vacuum, the imaginary part of the MIM response at \si{\giga\hertz} frequency $\omega \approx 0$ is given in linear response by~\cite{Taige2023comsol} 
\begin{equation}
    \Im \mathrm{MIM} \propto \int_{\br,\br'} G_{\br_t,\br} \chi_{\omega, \br,\br'} G_{\br,\br_t}
    \propto \int_{\bq} \chi(\omega, \bq) \frac{e^{-2|\bq| d}}{|\bq|^2},
    \label{eq:MIM_linear_response}
\end{equation}
where $\br,\br'$ are coordinates within the sample plane, $\br_t$ is the tip location, $G_{\br_t,\br'}$ is the (classical) electromagnetic Green's function in vacuum, and $\chi(\omega, \bq) = \int_{\br} e^{-i\bq (\br-\br')} \chi_{\omega, \br,\br'}$ is the real part of the charge susceptibility. Eq.~\eqref{eq:MIM_linear_response} assumes a homogeoneous (translation symmetric) sample; in the presence of disorder the signal will depend on the tip location $\br_t$, not just its vertical component $d$. Physically, this process comes from microwaves scattering into the sample and creating a charge density fluctuation. Those fluctuations create outgoing near-field microwaves, which scatter back to the tip and are measured.
For typical tip-sample distances, where $d$ is much larger than the unit cell scale, the leading order contribution to the MIM response is the charge compressibility: $\mathrm{MIM}\sim{} -i \frac{d\mu}{dn}$~\cite{Taige2023comsol}.
\\

\textbf{{Semiclassical model of the MIM response}}. 
For the finite element analysis (FEA) simulations, COMSOL Multiphysics is used for the frequency domain study. Specifically, the electric current (ec) module is used. The goal is to solve for the complex admittance of the tip-sample system at GHz frequencies to model the MIM response. Therefore, the electro-static solution and the corresponding boundary conditions are no longer relevant. This allows the system to be defined in terms of the conductivity tensors, without having to specify the $V_{\text{TG}}$ and $V_{\text{BG}}$ in the electrostatic limit.

In our 3D model, the two graphene layers are defined to be $\SI{20}{nm}$ thick. The tip-sample distance is defined to be $\SI{88.2}{nm}$ which captures the real tip-sample distance plus the thickness of the capping hBN layer on the heterostructure. The tip is modeled with a diameter of $\SI{100}{nm}$, which is negligible compared to the dimension of the simulated graphene layers  ($\SI{20}{\mu m}\times\SI{12}{\mu m}$). This ensures that the signal that the tip picks up primarily arises from the bulk of the quantum Hall insulator, minimizing the edge state contributions. Accordingly, we only input the longitudinal conductivities $\sigma^{xx}_\mathrm{T}$ and $\sigma^{xx}_\mathrm{B}$ into the simulation. We define the lateral position of the tip to be at the center of the sample so that the transverse conductivity $\sigma^{xy}$ becomes irrelevant. By solving this system at RF frequency $\SI{4.22}{GHz}$ and evaluating the admittance which is proportional to the experimental MIM signal, we can get the MIM response corresponding to any combination of $V_{\text{TG}}$ and $V_{\text{BG}}$.

The simulation tries to reveal the conductivity profile in the dual-gate sweep that corresponds to the longitudinal conductivity tensor shown in Fig.[\ref{fig:LL}](a-b). The trajectories of the top-layer filling factors $\nu_{\text{T}} = \pm 2$ and $\nu_\text{T} = 0$ indicated by pink dotted lines can be clearly observed in both simulation and experimental results, including the kinks along the plateau corresponding to the compressible state of the bottom graphene.

\textbf{New capabilities: MIM and local compressibility measurements with displacement field control.} 

\begin{figure}[H]
\centering
\includegraphics[width=0.28\textwidth]{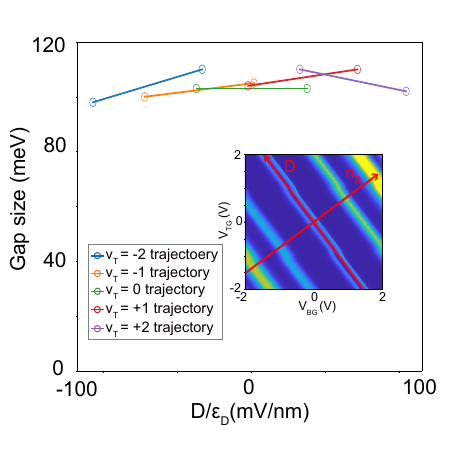}
\caption{\label{fig:Supp_Dfield} 
Gap sizes of the $\nu_{\text{B}} = \pm 2 $ state in the bottom graphene layer, measured as a function of vertical displacement field. 
\textit{Inset:} Calculated longitudinal conductivity, $\sigma^{xx}_\mathrm{B}$, of the bottom sample layer as a function of $V_{\text{TG}}$ and $V_{\text{BG}}$. The red axes label the vertical displacement field $D$ and the bottom layer carrier density $n_\text{B}$. }
\end{figure}

By using the ``sensor'' layer as a top gate, this double-layer imaging approach naturally provides a route for MIM and local electronic compressibility measurements with displacement field control. Because the sensor hosts a ladder of Landau levels that extend arbitrarily upward in energy, one can image the chemical potential landscape $\mu_\text{B}(x,y, n_\text{B})$ of the lower layer at a discrete sequence of energies corresponding to higher LLs in the top layer. This is equivalent to measuring $\mu_\text{B}(x,y, n_\text{B})$ at a series of displacement fields defined by the sequence of voltages $(V_\text{BG}^{\nu_\text{T} = n}, V_\text{TG}^{\nu_\text{T} = n})$ following the LL trajectories $\nu_\text{T} = n$ for $n=1, 2, 3 \dots$ in the $V_\text{BG} – V_\text{TG}$ plane.
As a proof of concept, we extracted the LL gaps of the sample layer, $\Delta\mu_\text{B} (\nu_{\text{B} }=\pm 2)$, as a function of the vertical displacement field. This is shown in Supp Fig.~\ref{fig:Supp_Dfield}: the inset shows the corresponding carrier density and displacement field axes in the bottom graphene layer, based on the modeling results presented in Fig.~\ref{fig:LL}(b). It should be noted that since the sample's gap can only be measured when the top graphene layer is in a Landau level gap, the displacement field is only achievable in a discrete sense, and thus cannot be tuned continuously at a fixed magnetic field. 
In monolayer graphene, the gap is not tunable by the vertical displacement field, which matches our findings, but this strategy can also be applied to other materials like multilayer graphene and twisted heterostructures, where exploration of the phase diagram requires transverse electric field control.
Additionally, the ability to image through top gates provides a route for probing  epitaxial semiconductors and topological materials that cannot be back gated, include cadmium arsenide \cite{rashidi2023Cd3As2,munyan2023Cd3As2} and magnetically doped (Bi,Sb)Te \cite{mogi2017BiSbTe,allen2019MIM}. 
This all-electronic readout method eliminates the need for complex optical measurement schemes, making this approach easily compatible with mK environments.

\newpage

\section*{Supplementary Figures}

\begin{figure}[H]
\centering
\includegraphics[width=0.42\textwidth]{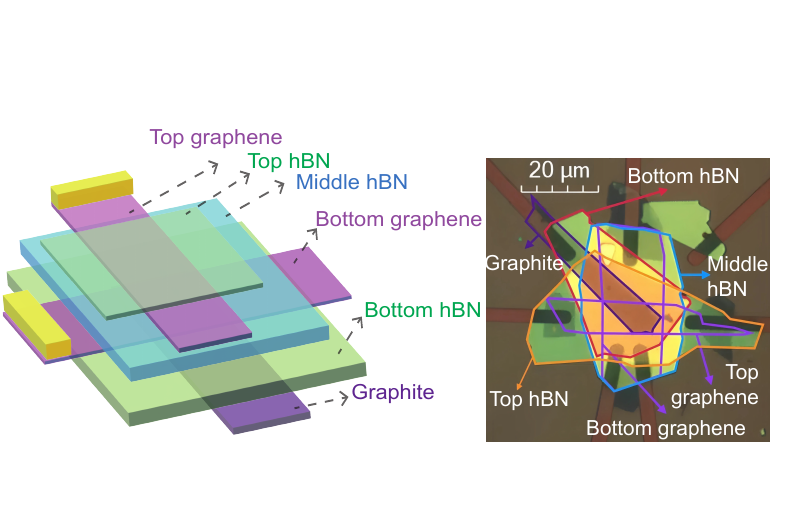}
\caption{\label{fig:Optical}Left: Schematics of the device layout, illustrating the actual alignment between individual layers in the fabricated hetero-structure. Right: Optical image of the device, with the different layers labeled in white.}
\end{figure}

\begin{figure}[H]
\centering
\includegraphics[width=0.42\textwidth]{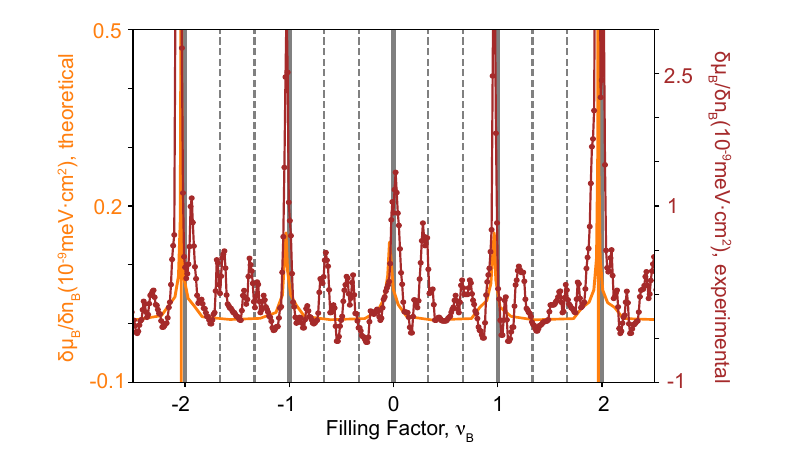}
\caption{\label{fig:Supp_inv_comp}Theoretical inverse compressibility (left, orange) and experimentally measured value (right, brown) plotted as a function of the bottom layer filling factor, $\nu_\text{B}$. The experimental value is extracted from Fig.~\ref{fig:LL}(f). }
\end{figure}

\end{appendix}

\end{document}